\documentclass[12pt]{spieman}  
\usepackage{amsmath,amsfonts,amssymb}
\usepackage{graphicx}
\usepackage{setspace}
\usepackage{tocloft}
\usepackage{tabularx}
\usepackage{longtable}
\usepackage{caption}
\usepackage{subcaption}

\title{A Predictive Model for Micrometeoroid Damage to Gossamer Structures}

\author[a]{Michaela N. Villarreal}
\author[a]{Jonathan W. Arenberg}
\author[a]{Lauren Halvonik Harris}

\affil[a]{Northrop Grumman Corporation, Redondo Beach CA, USA 90278}

\cftpagenumbersoff{figure}
\cftpagenumbersoff{table} 
\begin{document} 
\maketitle

\begin{abstract}
A typical inflatable reflector for space application consists of two thin membranes with a parabolic shape. It is critical to understand the interaction of the inflatable and the micrometeoroid environment to which it is exposed. This interaction leads to a series of penetrations of the inflatable membrane on entrance and exit of the impacting particle, creating a pathway for gas escape.  To increase the fidelity of the damage expected, we examine the literature for descriptions of micrometeoroid fragmentation and present a theoretical formulation for the damage caused by an impacting particle to the entrance and exit membranes. This theory is compared to an initial set of hyper-velocity tests for micrometeoroid-sized particles on thin film membranes. We use the results of these tests to produce a predictive model. This model is applied to estimate the damage rate near the 1 AU location and output predictions for the effectiveness of a micrometeoroid shield to reduce the damage on the lenticular and effectively optimize its lifetime. Lastly, we apply the kinetic theory of gasses to develop expressions for the expenditure of gas over a specified mission lifetime due to penetrations. Although this paper examines the specific case of an inflated lenticular protected by extra membrane layers, our predictive model can be applied for any gossamer structure composed of polyimide membranes.
\end{abstract}


{\noindent \footnotesize\textbf{*}Corresponding Author,  \linkable{jon.arenberg@ngc.com} }

\begin{spacing}{1} 

\section{Introduction}
\label{sect: intro}  

An inflatable reflector is typically composed of two thin sheets (such as Kapton) into a balloon-like structure with a parabolic shape. One side of the structure is aluminized to function as the primary reflector, while the other side is transparent to allow the incoming light to pass. The structure is supported by inflatant which enable the sheets to take on the proper shape. In 1996, the Inflatable Antenna Experiment (IAE) was carried out, proving that such a design can be successfully deployed (Freeland et al., 1997)\cite{Freeland1997}.  

Inflatable reflectors offer a direct path to unlock large ($>$ 10 $m$ diameter) aperture space telescopes that would not be implementable with a rigid mirror design. These large apertures can significantly advance the current understanding of astrophysics and planetary systems by increasing the signal to noise ratio and the number of targets that can be observed in a given time frame.\cite{ArenbergArch22} Hence, maximizing the lifetime of these systems is highly desired. 

Inflatable systems have one main vulnerability which limit their lifetimes: micrometeoroid impacts from the space environment create holes in the structure which allow the inflatant to escape. This requires replenishment of the gas to keep the lenticular properly inflated, the mass requirement for which sharply rises with lifetime (see Section \ref{sect: gas_loss_deriv}). Therefore, it is crucial to properly account for how quickly the space environment will create perforations in the inflatable structure.

A previous set of experiments were performed to analyze shape changes in response to pressure change, a thermal gradient, and a controlled puncture for a 1 $m$ inflatable optic (Quach et al., 2021)\cite{Quach21}. The controlled puncture experiment used a needle with a diameter of 600 $\mu m$ to simulate a micrometeoroid impact on one side of the reflector. No evidence of tear propagation was observed. The reflector exhibited fluctuations in its surface shape for tens of minutes before re-stabilizing. Note that the size of the needle diameter corresponded to a fairly large-sized micrometeoroid (see Section \ref{sect: mitigation}, and therefore the response represents a near worst-case scenario.  

The purpose of this paper is to create a predictive model for the damage incurred on an inflatable optic over time due to its local micrometeoroid environment. In Section \ref{sect:Theoretical_section} we review the theory for grain fragmentation and present a formulation for how to calculate the total damage from an impacting micrometeoroid. Section \ref{sect: hypervelocity_tests} covers results from new hypervelocity tests conducted at the White Sands Test Facility to mimic such events. These results are modeled using the formulation from Section \ref{sect:Theoretical_section} to determine the empirical values for the relevant parameters and construct a predictive model. This model is then applied to estimate the requirements to compose a structure capable of sufficiently breaking down incoming micrometeoroids so that little to no fragments impact the lenticular. Finally, we identify the gas flow regime for escape through impact holes and derive an expression for the total gas required to replenish the lenticular over a specified mission lifetime.

\section{Theoretical Formulation for Micrometeoroid Punctures and Fragmentation}
\label{sect:Theoretical_section}

In this section, we will review the fundamental physics involved in micrometeoroid impacts.

\subsection{Micrometeoroid Puncture Size on Entrance and Amplification of Damage}
\label{sect: mmod_entrance}

\begin{figure}
\begin{center}
\begin{tabular}{c}
\includegraphics[width=12cm]{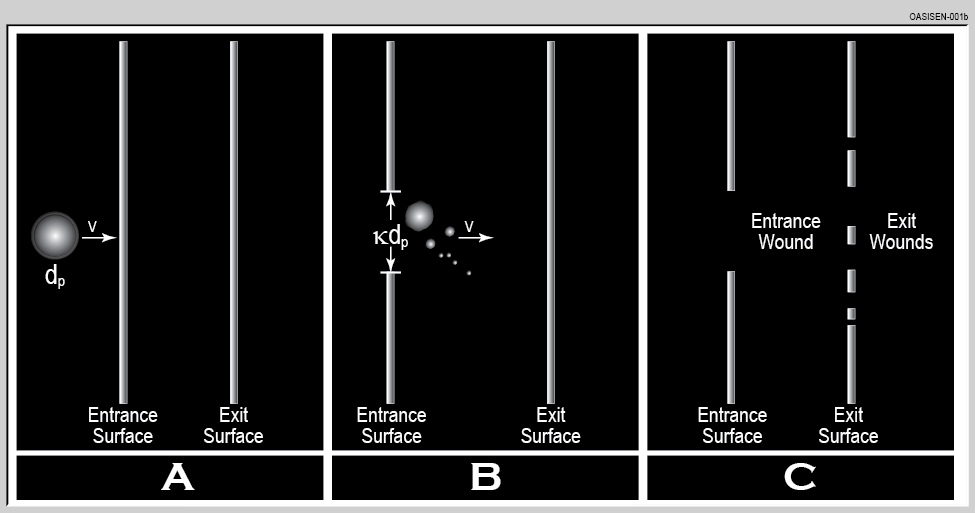}
\end{tabular}
\end{center}
\caption 
{\label{fig:frag_diagram}
Schematic of the stages of a micrometeoroid impact. Panel A: Incoming micrometeoroid impacts the first membrane, creating a single hole. Panel B: Micrometeoroid shatters after impact, creating a distribution of fragments that impact the second membrane. Panel C: Resulting damage allows a mechanism for gas escape.} 
\end{figure}

 A single micrometeoroid will cause damage to both sides of the inflatable structure as it enters and exits. These damage areas will not be equal: the micrometeoroid will shatter after encountering the first surface, creating many fragments which will subsequently cause many punctures on the exit surface (see Figure \ref{fig:frag_diagram}). This section focuses on calculating the damage upon the initial surface. To accomplish this, we will first review past literature that established the relationship between impactors and punctures. We will then leverage this knowledge to create an expression to estimate the damage for a single impact.
 
To begin, we first want to understand how the hole size varies with the size of the impactor. Gardner et al. (1997) \cite{Gardner} used hypervelocity tests on thin films to determine an equation which describes how to retrieve an impactor's diameter using the puncture size it left behind. They showed that an impactor's diameter can be inferred from the observed hole diameter with the expression
\begin{equation}
    \label{Eq:hole_growth}
d_p^\prime = \eta \left(\frac{10}{9+exp(D_h^\prime/\beta)}\right) + D_h^\prime \left(1-exp(D_h^\prime/\beta)\right)
\end{equation}
where $d_p^\prime=d_p/f$ is the ratio of the projectile diameter $d_p$ to the thickness of the film $f$, $D_h^\prime=D_h/f$ is the ratio of the hole diameter $D_h$ to the thickness of the film, and $\eta$ and $\beta$ are coefficients. The coefficient $\eta$ can be calculated by
\begin{equation}
 \label{Eq:A_coeff}
    \eta=6.97 \left(\frac{V\rho_p}{\sqrt{\sigma_t \rho_t}}\right)^{-0.723} \left(\frac{\sigma_t}{\sigma_{Al}}\right)^{-0.217} f^{-0.053}
\end{equation}
where $V$ represents the impactor's velocity, $\rho_p$ is the density of the projectile, $\rho_t$ is the density of the target, $\sigma_t$ is the Yield Stress of the target, $\sigma_{Al}$ is the Yield Stress of Aluminum, 6.90x10$^7$ $Pa$, and $f$ is the thickness of the film in $\mu m$. This equation can be applied using any set of internally consistent units, with the exception that $f$ and $d_p$ must be in units of $\mu m$. Note that the $\eta$ coefficient brings in the implicit dependencies of the hole size on the projectiles properties, such as density and velocity, as well as on the properties of the membrane, such as its density, thickness, and strength.

The coefficient $\beta$ is dependent on the impactor's velocity and behaves such that
\begin{equation}
 \label{Eq:B_coeff}
    \beta=\beta_1 + \beta_2 V
\end{equation}
where $\beta_1$ and $\beta_2$ are determined empirically for a specific target composition using hypervelocity data with $V$ in units of $km/s$.

\begin{figure}
\begin{center}
\begin{tabular}{c}
\includegraphics[height=7cm]{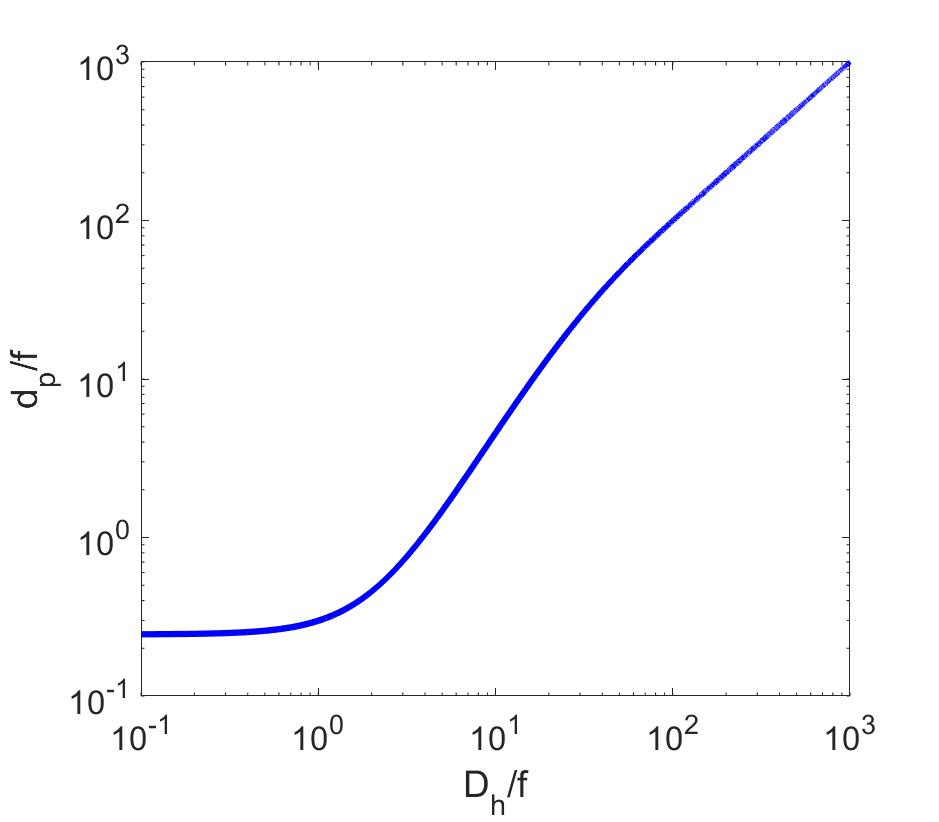}
\end{tabular}
\end{center}
\caption 
{ \label{fig:projectile_curve}
Equation \ref{Eq:hole_growth} is plotted for an impacting ruby with a velocity of 7 $km/s$ on 0.5 $mil$ Kapton film. This function will be used in Section \ref{sect: second_membrane_tests} to infer the diameters of the fragments, $d_f$ from observed holes. } 
\end{figure}

Figure \ref{fig:projectile_curve} shows the relationship between the projectile diameter $d_p$ and the hole radius observed $D_h$ for an impacting velocity of 7 $km/s$ and for input values representative of a ruby particle impacting a 0.5 $mil$ Kapton film (this will be the set-up for our experiments in Section \ref{sect: hypervelocity_tests}). The corresponding inputs are then $f=12.7$ $\mu m$, $\sigma_t=8.79x10^7$ $Pa$, $\rho_t=1380$ $kg/m^3$, and $\rho_p=3950$ $kg/m^3$.  The parameter $\beta$ is defined as $\beta=13.3+0.55V$ (see Section \ref{sect: first_membrane_tests} for more details). In Section \ref{sect: second_membrane_tests}, we will apply this function to infer the diameters of impacting fragments which are not known.

Gardner et al. (1997) \cite{Gardner} derived Equation \ref{Eq:hole_growth} specifically for metallic surfaces. Throughout this paper, we will apply this equation for the case of a Kapton (polyimide) membrane. Previous hypervelocity tests on Kapton have shown that it follows a similar overall trend to that derived by Gardner et al. (1997), but deviates in shape somewhat in the region between $D_h/f\sim 0.5-4$ (Neish and Kibe, 2001) \cite{Neish}. The purpose of future hypervelocity tests is to obtain enough data to derive a similar function specifically for polyimide films.

Equation \ref{Eq:hole_growth} is useful to back-out the particle diameter for an observed puncture. However, in this paper, we are interested in the reverse problem: given a particle diameter, what is the corresponding hole diameter? We introduce an amplification factor, $\kappa$, which describes the relationship between the particle diameter and the hole size it creates (for prescribed interaction conditions as defined in Equation \ref{Eq:hole_growth}) as
\begin{equation}
\label{Eq:kappa}
    \kappa=\frac{D_h}{d_p}.
\end{equation}
Equation \ref{Eq:hole_growth} can therefore be used to determine a $\kappa$ value for a corresponding incoming particle size $d_p$ given a set of input conditions. This involves first mapping $d_p$ to $D_h$ with Equation \ref{Eq:hole_growth}, then plugging in these respective values into Equation \ref{Eq:kappa}. Equation \ref{Eq:hole_growth} reveals that the amplification factor $\kappa$ at a specific $d_p$ is not constant, but varies with the properties of the impactor and the film, as well as the relative velocity between the two. 

Approximating the micrometeoroid as a spherical grain, the damage area caused to the first membrane (which we will refer to as the entrance membrane), $A_{ent}$, can be written as
\begin{align}
   A_{ent}= \frac{\pi}{\cos \theta} \left(\frac{D_h}{2}\right)^2  \\
A_{ent}= \frac{\pi}{\cos \theta} \left(\frac{\kappa d_p}{2} \right)^2  \\ 
\label{Eq:A_ent}
A_{ent}=\frac{\pi \kappa^2}{4 \cos \theta} d_p^2
\end{align}
where $\theta$ is the angle of incidence. Therefore, the resulting hole is an ellipse that depends on the angle of incidence. 

Gardner et al. (1997) \cite{Gardner} also showed that the minimum particle size required to be above the ballistic limit for a given velocity $V$ can be found by
\begin{equation}
\label{Eq:ballistic_limit}
    d_{p,bal}=\left(\frac{f}{0.129\left(\frac{V\rho_p}{\sqrt{\sigma_t \rho_t}}\right)^{0.763}\left(\frac{\sigma_t}{\sigma_{Al}}\right)^{0.229}}\right)^{1/1.056}
\end{equation}
where all terms maintain the same definitions as before. Similar to  Equation \ref{Eq:A_coeff}, this equation is valid as long as all variables are in a consistent set of units, with the exception that $f$ and $d_{p,bal}$ must be given in $\mu m$. We will utilize this equation later to determine which micrometeoroids are capable of penetrating the membrane.

\subsection{Micrometeoroid Fragmentation and Puncture Size on Exit}
\label{sect: mmod_exit}

Micrometeoroids near 1 AU generally impact with velocities of 35 $km/s$ or less (Thorpe et al., 2016\cite{Thorpe2016}). At these impact speeds, the micrometeoroid incident on the entrance membrane will shatter (spalls) into a number of smaller fragments. This collection of fragments will then produce many holes on the exit surface. In this section, we will devise an expression for the particle distribution as a result of this fragmentation. We then determine the total damage on the exit membrane due to secondary impacts from these fragments.

We follow the literature for the shattering of interstellar dust grains \cite{MRN}\cite{Jones}\cite{hirashita} (Mathis et al., 1997; Jones et al., 1996;  Hirashita and Kobayashi, 2013) to describe the fragment distribution after the micrometeoroid experiences catastrophic destruction from encountering the first surface. It has been shown that a dust grain that undergoes catastrophic destruction will shatter into smaller pieces with a number distribution that can be described by a power law \cite{MRN}\cite{Jones}
\begin{equation}
\label{Eq:nm_distribution}
    n_f (m_f)dm_f=c_f \left(\frac{3m_f}{4 \pi \rho}\right )^{-\frac{\alpha_f}{3}} dm_f
\end{equation}
where the number distribution of the fragments is denoted by $n_f$ over an interval $(m_f,m_f+dm_f)$ with $m_f$ the mass of the fragment, $c_f$ is a scaling coefficient, $\rho$ is the mass density, and $\alpha_f$ is the scaling exponent. We will assume the impactor has a uniform density, so that the fragments maintain the same density as the original projectile. The literature suggests $\alpha_f$ is in the range from 3-3.5 \cite{MRN}\cite{Jones}\cite{hirashita}.

The number distribution will be bounded by a minimum and maximum fragment size. The maximum fragment size is believed to be proportional to the projectile radius, $a_p$, such that \cite{Jones}
\begin{equation}
\label{Eq:a_max}
    a_{f,max}=c_{max} a_p
\end{equation}
where $a_{f,max}$ is the radius of the largest fragment and  $c_{max}$ is the scaling coefficient. For the case of catastrophic destruction of the impactor, Jones et al. (1996) \cite{Jones} adopt a value of $c_{max}$ =0.22, while Hirashita and Kobayashi (2013) \cite{hirashita} use a value of $c_{max}$ =0.27. Both applications assume a minimum fragment size $a_{f,min} \sim$ 5x$10^{-10}$ $m$ as the shattering limit.

We can determine $c_f$ by invoking the conservation of mass. The total mass of all the fragments must equal the mass of the original particle $m_p$ and thus
\begin{equation}
\label{Eq:conservation_mass}
    m_p= \int_{m_{f,min}}^{m_{f,max}} n_f m_f dm_f.
\end{equation}
Substituting Equation \ref{Eq:nm_distribution} into Equation \ref{Eq:conservation_mass} gives
\begin{equation}
    m_p= \int_{m_{f,min}}^{m_{f,max}} c_f \left( \frac{3m_f}{4\pi \rho} \right)^{-\frac{\alpha_f}{3}}  m_f dm_f.
\end{equation}
This simplifies to
\begin{equation}
    m_p= c_f \left(\frac{4 \pi \rho}{3} \right)^{\frac{\alpha_f}{3}} \int_{m_{f,min}}^{m_{f,max}} m_f^{1-\frac{\alpha_f}{3}} dm_f.
\end{equation}
This integrates to
\begin{equation}
    m_p=\left(\frac{4 \pi \rho}{3} \right)^{\frac{\alpha_f}{3}} \frac{c_f}{2-\frac{\alpha_f}{3}} \left(m_{f,max}^{2-\frac{\alpha_f}{3}} - m_{f,min}^{2-\frac{\alpha_f}{3}} \right).
\end{equation}
We can rewrite the above equation in terms of the original impactor radius, $a_p$, and the minimum and maximum fragment radii by using $m=\frac{4\pi}{3} \rho a^3$, which results in
\begin{equation}
    \frac{4\pi}{3} \rho a_p^3 = \left(\frac{4 \pi \rho}{3} \right)^2 \frac{c_f}{2-\frac{\alpha_f}{3}} \left(a_{f,max}^{6-\alpha_f} - a_{f,min}^{6-\alpha_f} \right).
\end{equation}
This simplifies to
\begin{equation}
  a_p^3 = \frac{4 \pi \rho c_f}{6-\alpha_f} \left(a_{f,max}^{6-\alpha_f} - a_{f,min}^{6-\alpha_f} \right).
\end{equation}
This can be rearranged to solve for $c_f$ as
\begin{equation}
 \label{Eq:C_f}
    c_f= \frac{6-\alpha_f}{4 \pi \rho } \frac{a_p^3}{a_{f,max}^{6-\alpha_f} - a_{f,min}^{6-\alpha_f}}.
\end{equation}
From Equation \ref{Eq:C_f} we can see that the $c_f$ coefficient will depend on the impacting particle's radius $a_p$, the particle's density $\rho$, proportionality constant $c_{max}$ (through the dependence on $a_{f,max}$), the minimum fragment size $a_{f,min}$, and the scaling exponent $\alpha_f$.

The total damage area incurred by the distribution of fragments can be estimated by characterizing each individual puncture according to Equation \ref{Eq:A_ent}. The total area of all the holes upon the exit surface is then expressed as
\begin{equation}
    \label{Eq:A_exit}
A_{exit}= \frac{\pi}{ \cos \theta}  \int_{m_{f,min}}^{m_{f,max}} n_f \kappa^2 a_f^2 dm_f.
\end{equation}
 For simplicity, we assume all the fragments retain the same angle of incidence $\theta$ as the original impactor and thus is taken outside of the integral. Since the $\kappa$ term is ultimately dependent on $a_f$, and therefore $m_f$, it is kept in the integral and will need to be varied for each fragment size according to the implicit dependencies within Equation \ref{Eq:hole_growth}. Equation \ref{Eq:A_exit} is simple to compute numerically.
 
 We embarked on this formulation to ultimately determine the total damage that a single micrometeoroid will cause to both sides of the lenticular, which we will call $A_{impact}$. To estimate the total damage area in the absence of a micrometeoroid shield, we can define $A_{impact}$ as
 \begin{equation}
 \label{Eq:A_impact}
     A_{impact}=A_{ent}+A_{exit}
 \end{equation}
where $A_{ent}$ and $A_{exit}$ are defined by Equations \ref{Eq:A_ent} and \ref{Eq:A_exit}. It should be noted that our model assumed that the membrane material is vaporized on impact and that penetrations made on later layers are the result of the repeated fragmentation of the original impactor.

In Section \ref{sect: gas_loss_deriv}, we will want to determine the gas lost through these punctures over time. The cumulative hole area, $A_H$, in the primary reflector can be estimated by 
\begin{equation}
    \label{Eq42}
    A_H(t)=A_{seams}+A_{R}t \int_{m_{p,min}}^{m_{p,max}} \Phi(m_p) A_{impact} dm_p
\end{equation}
where $A_{seams}$ is the initial hole area due to construction of the seams, $A_{R}$ is the surface area of the primary reflector, $t$ is time, $\Phi (m_p)$ represents the micrometeoroid flux as a function of projectile mass $m_p$, and $A_{impact}$ describes the hole area caused under impact conditions for the micrometeoroid's radius $a_p$ (see Equations \ref{Eq:hole_growth} and \ref{Eq:A_impact} for all inherent dependencies). The $\Phi(m_p)$ term is a property of the local space environment, where the bounds $m_{p,min}$ and $m_{p,max}$ represent the minimum and maximum incoming micrometeoroid sizes.

\subsection{Micrometeoroid Punctures after N Shattering Events}
\label{sect: mmod_n_layers}

In Section \ref{sect: mitigation}, we will discuss possible mitigation of micrometeoroid holes through the use of a micrometeoroid shield, which consists of multiple layers of Kapton that act as a bumper for the lenticular. The total fragment mass on each consecutive interface should decrease for two reasons: vaporization and the inability of the smallest fragments to reach the ballistic limit. The goal is to fragment the micrometeoroid sufficiently so that only a few fragments remain above the ballistic limit when they arrive at the lenticular structure at the center. This will decrease the hole area with time ($A_H(t)$ in Equation \ref{Eq42}) and  allow inflatant to be conserved. Since the fraction of mass that vaporizes is difficult to quantify with experiments, this paper will focus on the mass loss due to the ballistic limit.

We set-up the formulation needed to propagate the fragmentation for consecutive membrane layers to determine the damage incurred on each interface. We can think of repeated fragmentation as a fractal process: each fragment can be thought of as the original impactor  described in Section \ref{sect: mmod_exit}, and Equations \ref{Eq:nm_distribution}, \ref{Eq:a_max}, and \ref{Eq:C_f}  can be evaluated in the same manner as Section \ref{sect: mmod_exit} to determine the secondary fragmentation of each incoming fragment.

We assume that the fragments, similar to the original projectile, further fragment according to the powerlaw distribution in Equation \ref{Eq:nm_distribution}, with each fragment having a maximum secondary fragment size of
\begin{equation}
a_{f,max,s}=c_{max}a_{f}
\end{equation}
where the $s$ subscript represents a secondary fragment. This means that the largest fragment out of all the secondary fragments after the N$^{th}$ shatter can be calculated by
\begin{equation}
    \label{Eq:a_fmax_sec}
a_{f,max,N}=c_{max}^N a_p.
\end{equation}

We will again assume the fragments shatter as spheres, allowing us to define the maximum fragment mass as $m_{f,max,N}=\frac{4\pi}{3} \rho a_{f,max,N}^3$. For simplicity, we assume all subsequent fragmentations maintain the same $a_{f,min}$ as the original impactor. We also assume that fragments will not fragment further once they reach a size of $a_{f,min}$. The number distribution of fragments at size $m_f$ after the N$^{th}$ shattering event, $n_{f,N}$, can then be calculated by

\begin{equation}
    \label{Eq:nm_N_layers}
n_{f,N}(m_f)dm_f=\sum_{i=m_{f,min}}^{m_{f,max,N-1}} q_i n_{i,s}   
\end{equation}
where $q$ is the number of incoming fragments at size $i$ from the previous shattering, expressed by
\begin{equation}
    q_i=n_{f,N-1}(i)dm_f 
\end{equation}
and $n_{i,s}$ represents the number of secondary fragments of size $m_f$ produced by the shattering of the impacting fragment of size $i$, which can be written as
\begin{equation}
   n_{i,s}(m_f)dm_f= c_{i,s} \left(\frac{3m_{f}}{4 \pi \rho}\right )^{-\frac{\alpha_f}{3}} dm_{f}.
\end{equation}
 The scaling term $c_{i,s}$ will need to be re-calculated for each incoming fragment size $i$ being evaluated as discussed in Section \ref{sect: mmod_exit} (here, each impacting fragment $m_i$ is now treated as the $m_p$). The number distribution for $n_{f,N}$ will be bounded from $m_{f,min}$ to $m_{f,max,N}$, as defined earlier. This fragmentation process will repeat until all the fragments reach a size of $a_{f,min}$ and the particles cannot shatter any further. Although Equation \ref{Eq:nm_N_layers} appears cumbersome, is it relatively easy to implement numerically.

The total damage area incurred on the membrane after the N$^{th}$ shatter is thus
\begin{equation}
    \label{Eq: A_damage_N_layers}
    A_N=\frac{\pi}{ \cos \theta}  \int_{m_{f,min}}^{m_{f,max,N}} n_{f,N} \kappa^2 a_f^2 dm_f.
\end{equation}
We again remind the reader that $\kappa$ is not a constant, but will vary as a function of the the micrometeoroid properties, film properties, and the impacting velocities.

\section{Laboratory Analysis of High Velocity Impacts on Membranes}
\label{sect: hypervelocity_tests}

In this section, we use hypervelocity impact tests on Kapton membranes to empirically define the parameter values for the micrometeoroid fragmentation from Section \ref{sect:Theoretical_section}.

\subsection{Experimental Design and Damage Area Observed}
\label{sect: experimental_design}

The purpose of these experiments is to aid us in the characterization of hole area damage for the micrometeroid environment of concern, as well as to provide data to support and validate our theoretical model. In particular, for the characterization of an the impact of a micrometeoroid through an initial wall, the subsequent break-up and debris cloud, and lastly that break-up and cloud’s impact upon the second wall. 

All testing was performed at NASA White Sands Test Facility (WSTF). A shot matrix was developed with the purpose of characterizing hole area as a function of the projectile’s velocity, the two target surface’s stand-off distance, and the impact angle. The test shot matrix is shown in Table \ref{tab:experiment_setup}. Due to testing restraints, we were not able to test a sampling of projectile diameters and mass. 

\begin{table}[ht]
\caption{Experimental design for each shot performed.} 
\label{tab:experiment_setup}
\begin{center}       
\begin{tabular}{|m{0.7cm}||m{1.8cm}||m{1.8cm}||m{2cm}||m{1.4cm}||m{1.5cm}||m{1.5cm}|}
\hline
\rule [-1ex]{0pt}{3.0ex}  Test $\#$ & Projectile Material & Projectile Diameter ($\mu m$) & Calculated Projectile Mass ($g$) & Impact Angle ($deg$) & Velocity ($km/s$) & Stand-off Distance ($m$) \\
\hline\hline
\rule[-1ex]{0pt}{3.0ex}  1 & Ruby & 200 & 1.65x10$^{-5}$ & 0 & 7.09 & 0.33 \\
\hline
\rule[-1ex]{0pt}{3.0ex} 2 & Ruby & 200 & 1.65x10$^{-5}$ & 0 & 7.05 & 0.66 \\
\hline
\rule[-1ex]{0pt}{3.0ex}  3 & Ruby & 200 & 1.65x10$^{-5}$ & 0 & 7.18 & 1.00 \\
\hline
\rule[-1ex]{0pt}{3.0ex} 4 & Ruby & 200 & 1.65x10$^{-5}$ & 50 & 7.18 & 0.33 \\
\hline
\rule[-1ex]{0pt}{3.0ex}  5 & Ruby & 200 & 1.65x10$^{-5}$ & 50 & 6.9 & 1.00 \\
\hline
\rule[-1ex]{0pt}{3.0ex}  6 & Ruby & 200 & 1.65x10$^{-5}$ & 0 & 4.93 & 0.33 \\
\hline
\rule[-1ex]{0pt}{3.0ex}  7 & Ruby & 200 & 1.65x10$^{-5}$ & 0 & 4.17 & 0.33 \\
\hline
\end{tabular}
\end{center}
\end{table} 

So as to accurately model the micrometeoroid environment, a 200 $\mu m$ diameter Aluminum Oxide (Al$_2$O$_3$, Ruby, Density 3.95 $g/cm^3$) projectile is used, decided upon with conference of the subject matter experts at WSTF. Prior to each test, the projectile mass was measured using a Mettler Toledo XP56 Delta Range balance (± 0.001 $mg$), while the  projectile diameter was measured using a Geller microanalytical laboratory micro-ruler (±0.01 $mm$).

\begin{figure}
\begin{center}
\begin{tabular}{c}
\includegraphics[height=6.5cm]{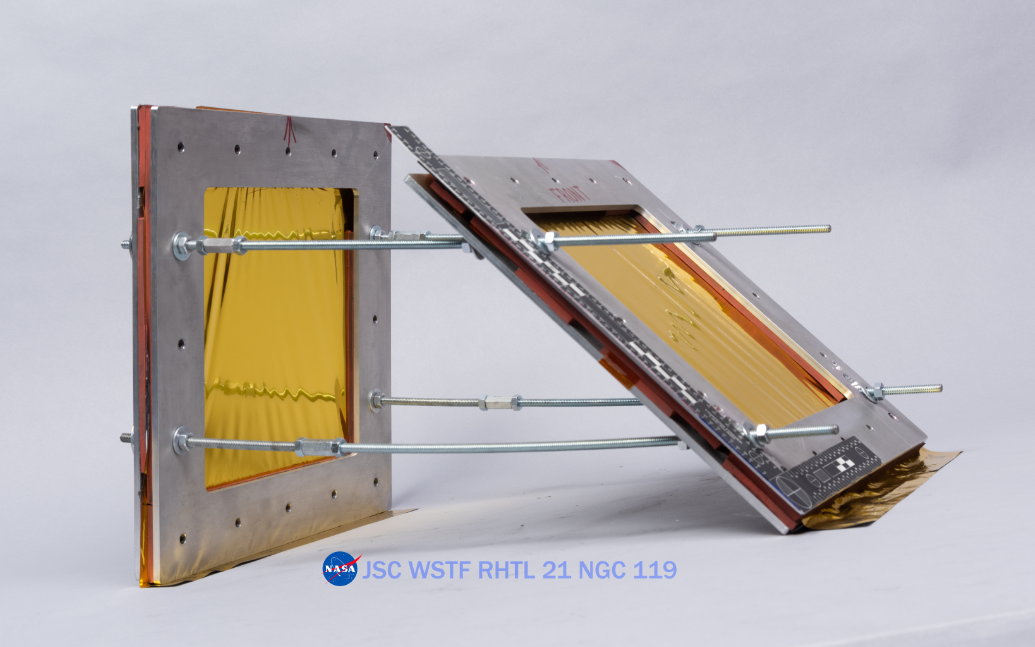}
\end{tabular}
\end{center}
\caption 
{\label{fig:test_setup}
Test Article Set-Up. The first membrane is set to the specified impact angle, while the second membrane surface is always set to a zero incidence angle. The stand-off distance between the membranes is varied for each test shot.} 
\end{figure}

The tests use kapton samples manufactured at Northrop Grumman Space Park. This is representative of the current choice of membrane for the OASIS mission concept whose design is based on an inflatable lenticular, and are also the same films that compose the sunshield for the James Webb Space Telescope. The target samples are kapton sheets of 0.5  $mil$ (12.7 $\mu m$) thickness, cut to approximately 12” square. The film samples are secured in frames as pictured in Figure \ref{fig:test_setup}, setting them to the appropriate incline angle and stand-off distance for a specified test shot (see Table \ref{tab:experiment_setup}). The samples were secured in manner such that they each experienced a tensile load of approximately 1 $ksi$. This load is reflective of the pressure that would be experienced by the membrane when fully inflated. The $\sim$1 $ksi$ stress level was achieved by attaching a small weight to the bottom part of the target membrane.  The first membrane is set to a specified incidence angle given in Table \ref{tab:experiment_setup}, while the second membrane remained at an incidence angle of zero for all test shots (see Figure \ref{fig:test_setup}). 

Seven impact tests were performed for a 200 $\mu m$ diameter ruby projectile upon the two consecutively placed 0.5 $mil$ (12.7 $\mu m$) thick Kapton membranes.  Images of the membranes were used to determine the damage area on each surface. Impact holes were identified by eye and the perimeters of the holes were defined.  The damage area in $pixels^2$ was calculated and subsequently converted to $mm^2$ using a scale bar on the image. Since there is a large number of holes on each second membrane due to fragmentation, a high density and low density damage region is defined. These regions are then divided into quadrants. A quadrant is selected to manually identify all the impact holes and the cumulative area is multiplied by four to calculate the total damage in that region. Table \ref{tab:obs_damage} lists the total damage area observed on each membrane. 

\begin{figure}
\begin{center}
\begin{tabular}{c}
\includegraphics[height=5.5cm]{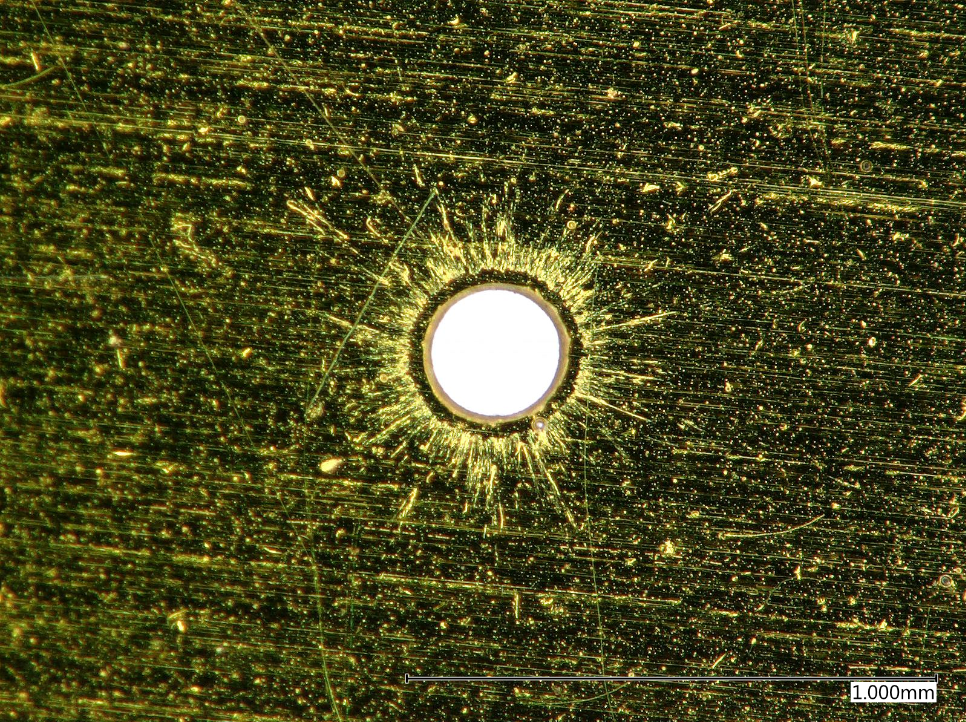}
\end{tabular}
\end{center}
\caption 
{\label{fig:6A}
Damage to the first membrane of Shot 6. A single, circular hole is observed.} 
\end{figure}

So as to further illuminate the image analysis method described above, let us look at the case of the sixth shot (Table \ref{tab:experiment_setup}). Figure \ref{fig:6A} shows the damage to the entrance membrane. The particle creates a well defined, circular hole which can directly be used to determine the $\kappa$ value for the particle's size and velocity (see Section \ref{sect: first_membrane_tests}). The other test shots also displayed circular impacts on the first membrane, with the exception of shots 4 and 5, which impacted at an incidence angle of 50 $deg$. The wounds from these impacts were elliptical in shape, consistent with Equation \ref{Eq:A_ent}. The total damage area to the entrance membrane, $A_{ent,obs}$, observed on each test shot is listed in Table \ref{tab:obs_damage}.

\begin{figure}
\begin{center}
\begin{tabular}{c}
\includegraphics[height=6cm]{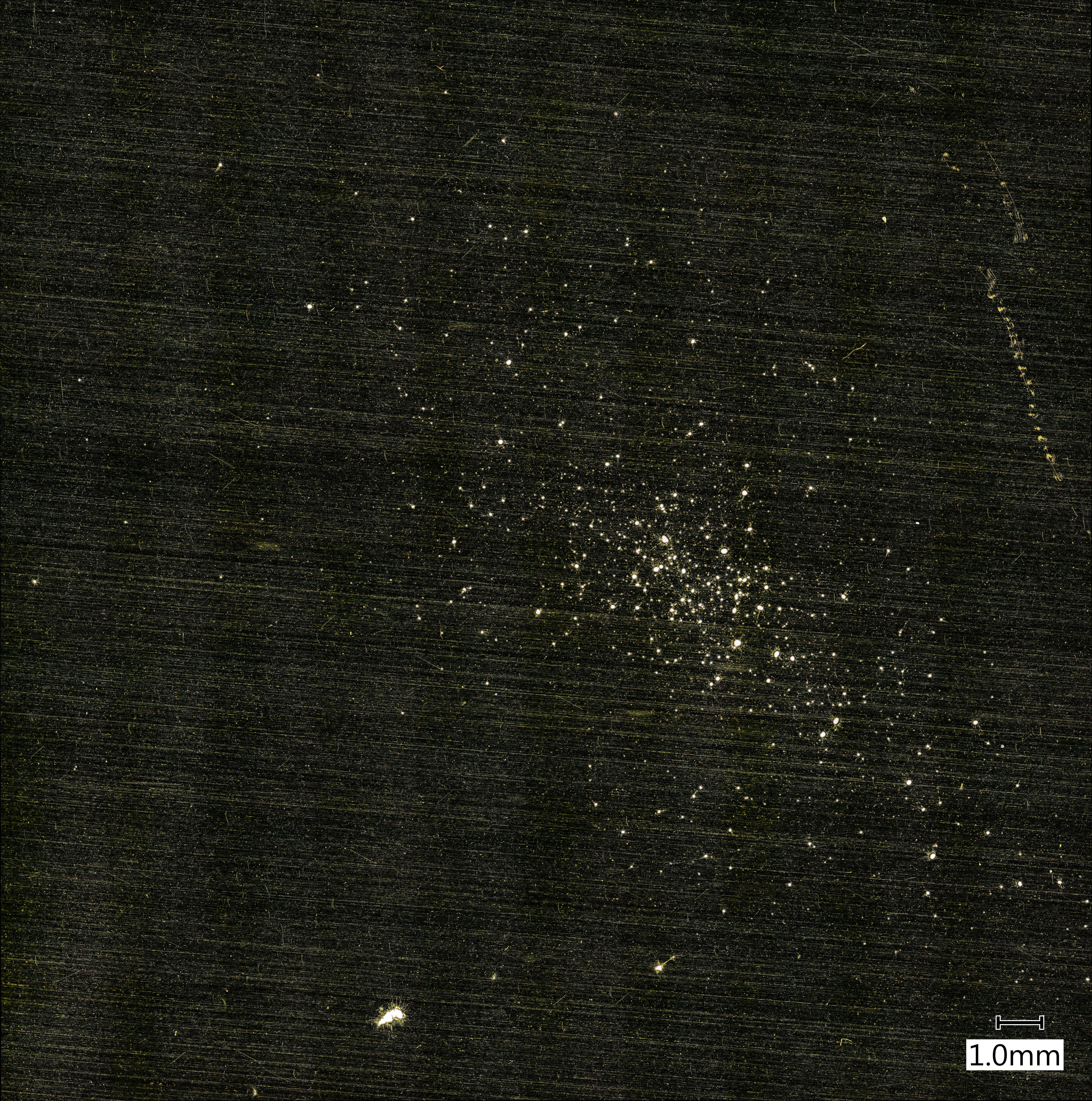}
\end{tabular}
\end{center}
\caption 
{\label{fig:6B}
Damage to the second layer of Shot 6. The impactor completely disrupts after impact with the first layer, resulting in fragments which create many holes on the second membrane. A high density damage region is observed in the center, surrounded by a diffuse region of damage.} 
\end{figure}

Figure \ref{fig:6B} shows the damage incurred on the exit membrane on shot 6. It is clear that the test particle has completely disrupted into many fragments, creating holes of various sizes. These fragments create a high density region of damage in the center, with more diffuse damage occurring further on the outer edges. 

\begin{figure}
     \centering
     \begin{subfigure}[b]{0.35\textwidth}
         \centering
         \includegraphics[width=\textwidth]{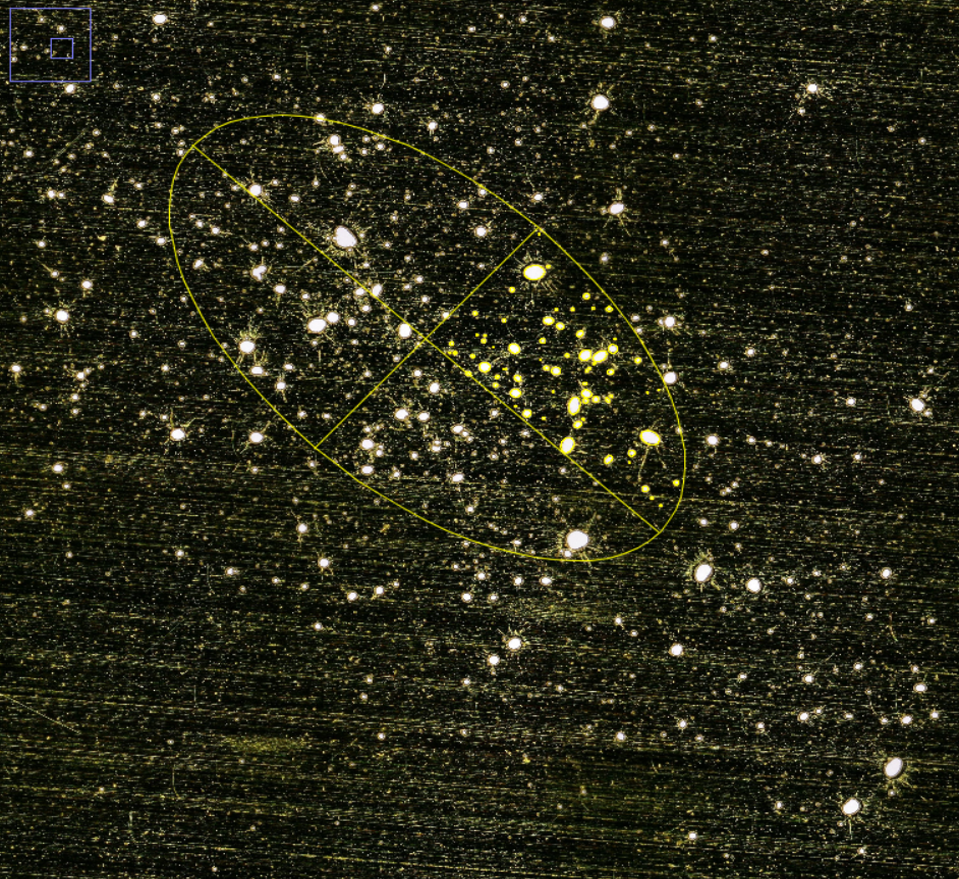}
     \end{subfigure}
     \hfill
     \begin{subfigure}[b]{0.35\textwidth}
         \centering
         \includegraphics[width=\textwidth]{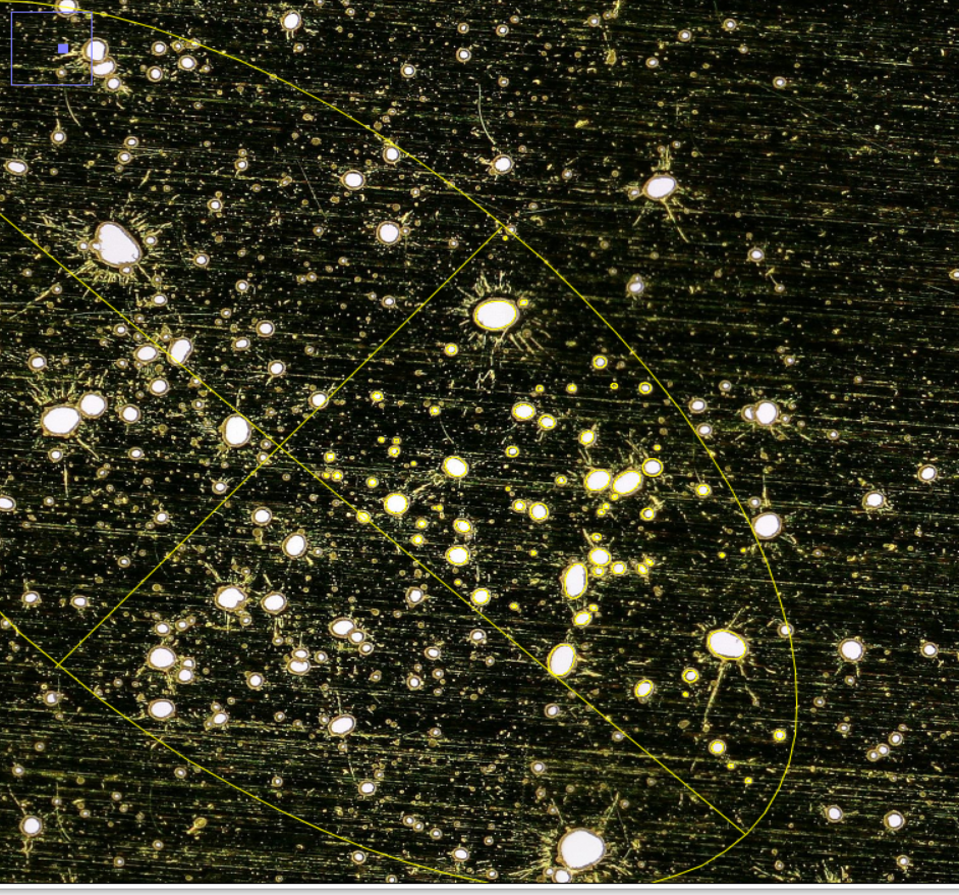}
     \end{subfigure}
        \caption{High density region on the second membrane of shot 6. $Left$: The high density region is defined and split into quadrants. A quadrant is selected (lower right) to manually measure each hole. $Right$: Close up of quadrant measurement.}
        \label{fig:6B_High_Density}
\end{figure}

\begin{figure}
\begin{center}
\begin{tabular}{c}
\includegraphics[height=7cm]{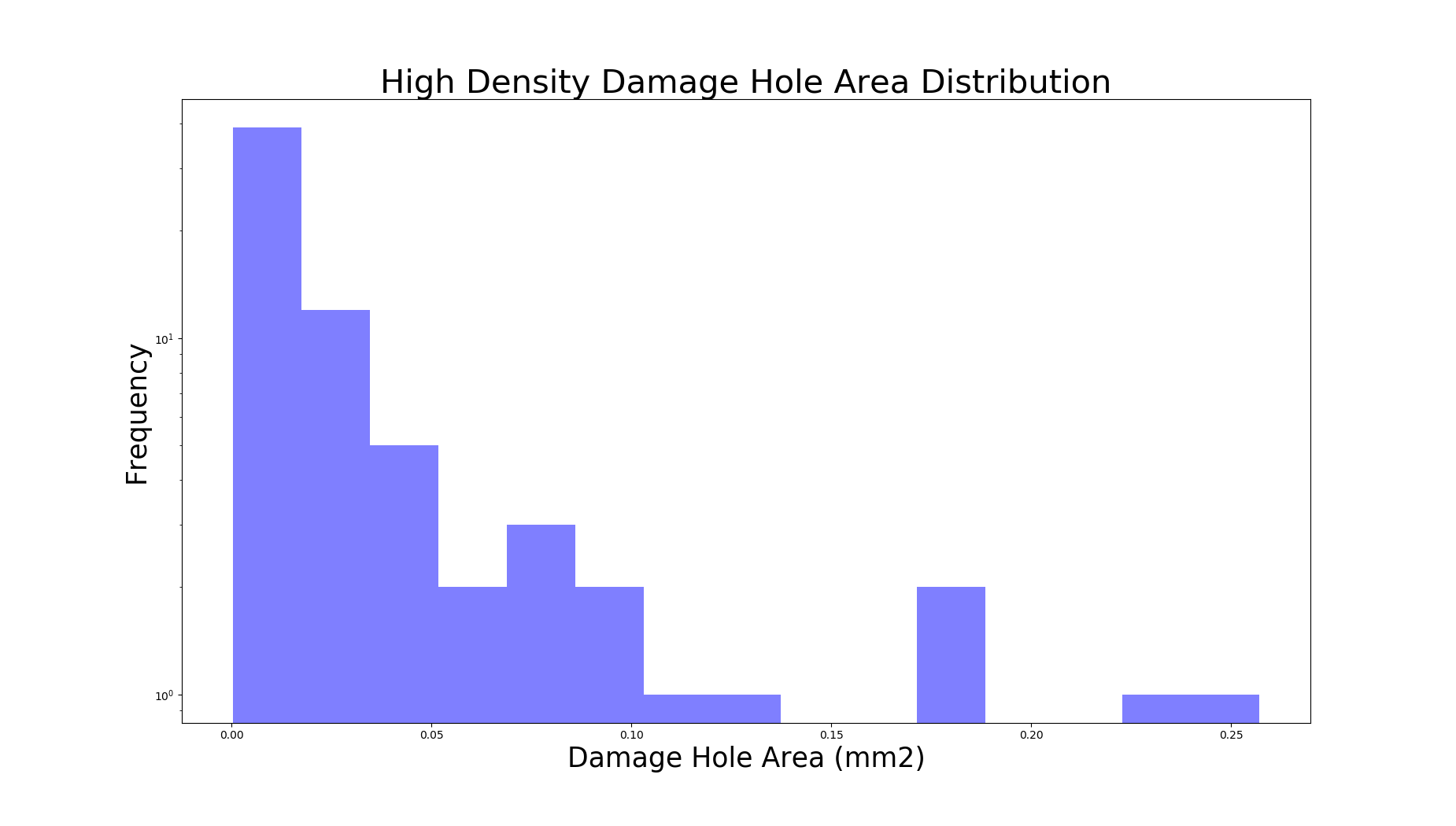}
\end{tabular}
\end{center}
\caption 
{\label{fig:6B_highdist}
The figure above depicts the size distribution of damage holes upon the back layer of shot 6 for the high density region. As can be seen, the damage area follows a power law distribution.} 
\end{figure}

The high density region of damage is identified in Figure \ref{fig:6B_High_Density}. The lower right quadrant is selected and every hole identified in that quadrant is measured, with measurements represented in the ImageJ software by a yellow circle. All measurements were exported to an excel document and then summed to get the full damage area of the quadrant. The quadrant area is multiplied by four to achieve the high density region's total damage area, $A_{High}$. The distribution of holes in this region is seen to follow a power law distribution as expected (Figure \ref{fig:6B_highdist}). 

\begin{figure}
     \centering
     \begin{subfigure}[b]{0.35\textwidth}
         \centering
         \includegraphics[width=\textwidth]{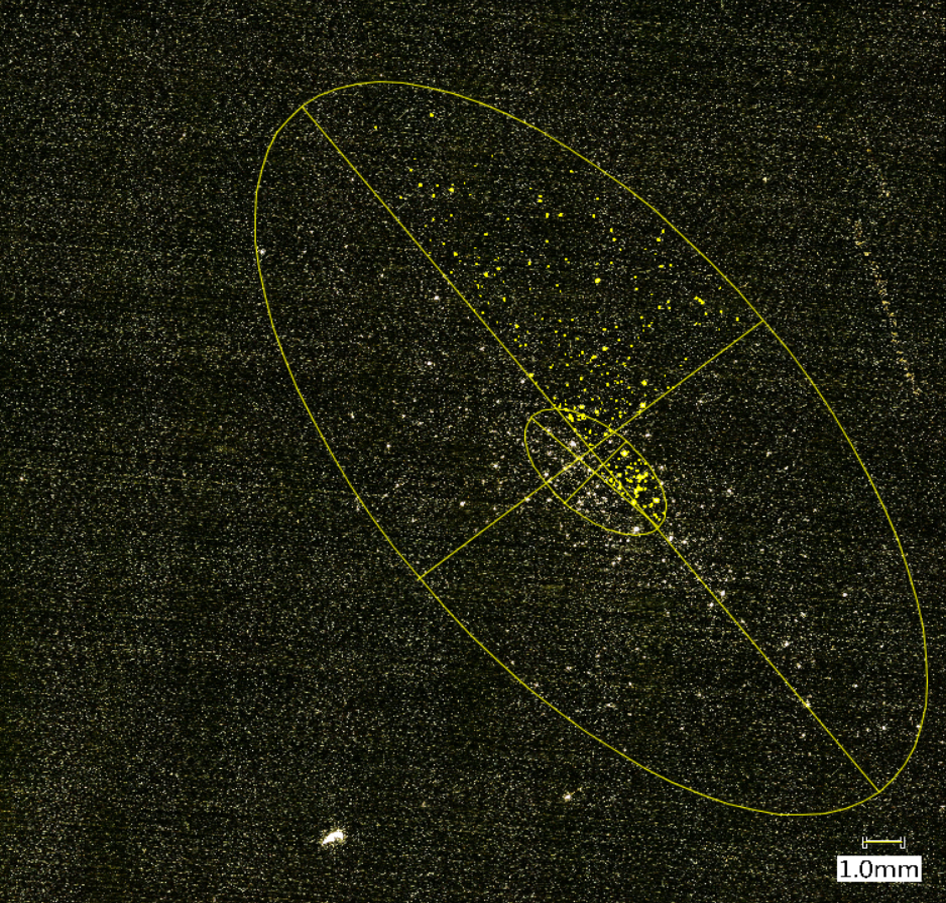}
     \end{subfigure}
     \hfill
     \begin{subfigure}[b]{0.35\textwidth}
         \centering
         \includegraphics[width=\textwidth]{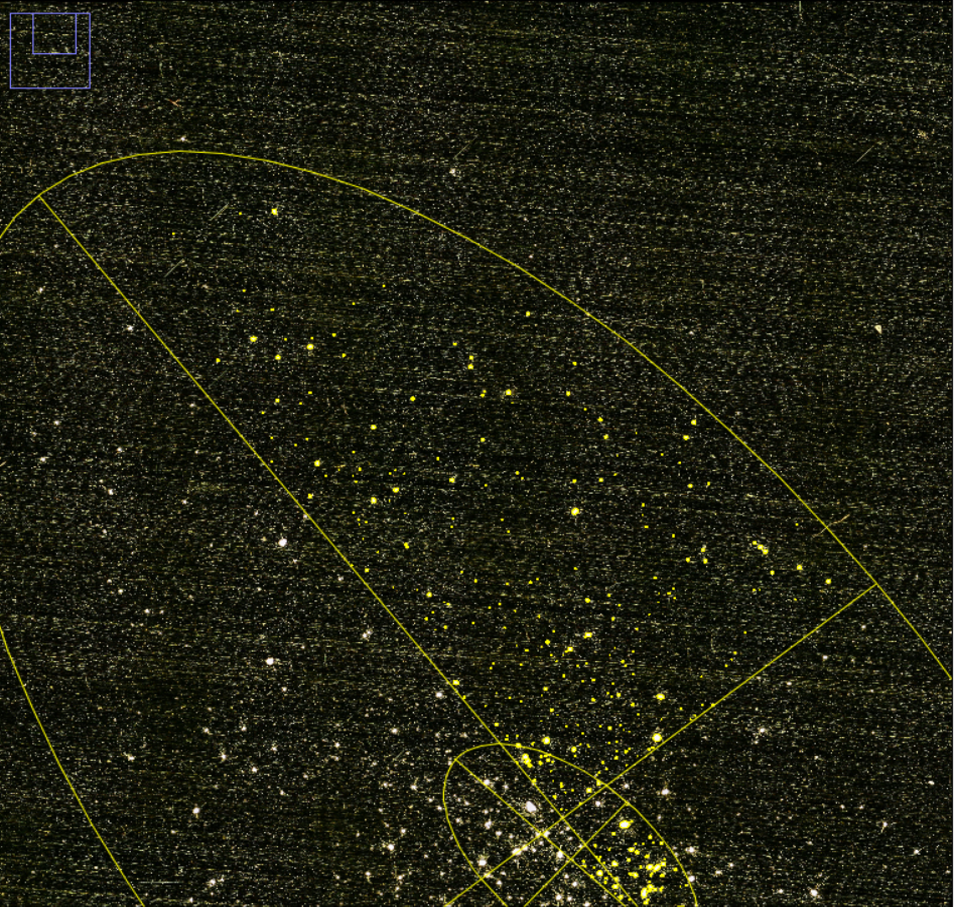}
     \end{subfigure}
        \caption{Low Density Region  on the second membrane of shot 6. $Left$: The low density region is defined and split into quadrants. A quadrant is selected (upper right) to manually measure each hole. $Right$: Close up of quadrant measurement.}
        \label{fig:6B_Low_Density}
\end{figure}

\begin{figure}
\begin{center}
\begin{tabular}{c}
\includegraphics[height=7cm]{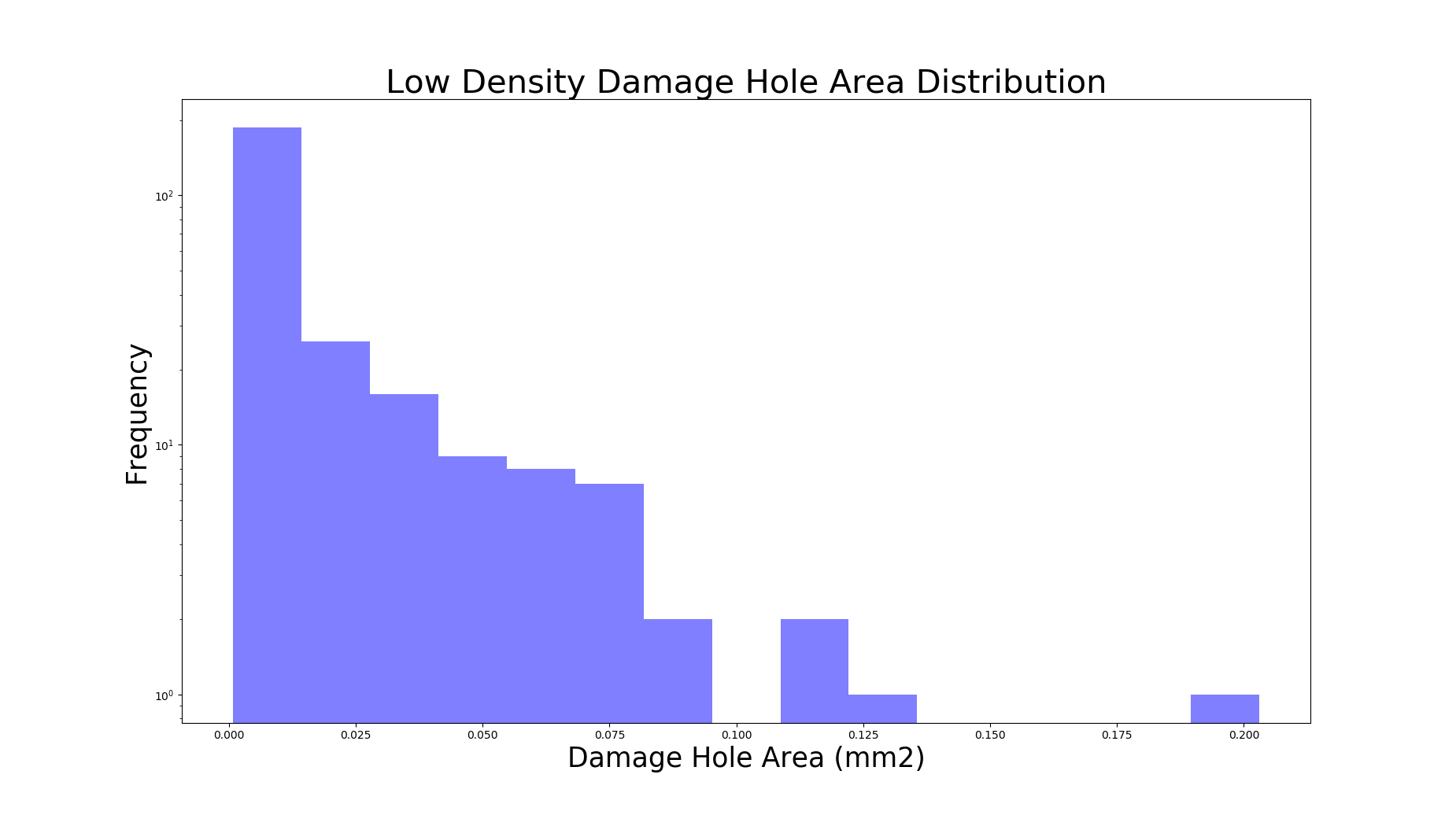}
\end{tabular}
\end{center}
\caption 
{\label{fig:6B_lowdist}
The figure above depicts the size distribution of damage holes upon the back layer of shot 6 for the low density region. As can be seen, the damage area follows a power law distribution.} 
\end{figure}

The low density region of damage is identified in Figure \ref{fig:6B_Low_Density}. The upper right quadrant is selected and every hole identified in that quadrant is measured, with measurements represented in the ImageJ software by a yellow circle. The distribution of holes in this region is seen to follow a power law distribution as well (Figure \ref{fig:6B_lowdist}). Compared to the high density region, the low density region has a higher percentage of smaller-sized holes, while the high density region contains a higher percentage of larger-sized holes. 

Since the high density region is contained within the low density region, we have to subtract the contribution of the high density region to the low density quadrant counts to avoid double counting. The total areal damage in the low density region, $A_{Low}$, is then calculated as
\begin{equation}
\label{Eq:count_correction}
    A_{Low}=4 (A_{q,Low}-\gamma A_{q,High})
\end{equation}
where $A_{q,Low}$ and $A_{q,High}$ is the total damage measured in the quadrants of the low density and high density regions respectively, and $\gamma$ represents the fractional area of the low density region occupied by the high density region. This quantity is multiplied by four to get the total damage over all four low density quadrants. The total damage to the exit membrane is then calculated by $A_{exit,obs}=A_{High} + A_{Low}$, listed in Table \ref{tab:obs_damage}.


All seven shots exhibit similar behavior for fragmentation and follow a power law distribution for hole sizes. Table \ref{tab:obs_damage} lists the total damage area observed on each membrane. Some variation in the damage distribution was observed with stand-off distance (i.e. high density and low density regions are less well defined for greater distances). However, it was difficult to determine the role of the stand-off distance in the total damage incurred since the velocity on each shot also changed. More data is needed to properly characterize this effect.

\begin{table}[ht]
\caption{Total damage observed on each membrane.} 
\label{tab:obs_damage}
\begin{center}       
\begin{tabular}{|m{1.5cm}||m{3.5cm}||m{3.5cm}|}
\hline
\rule[-1ex]{0pt}{3.5ex}  Test $\#$ & Total Damage to First Membrane ($\mu m^2$) & Total Damage to Second Membrane ($\mu m^2$) \\
\hline\hline
\rule[-1ex]{0pt}{3.5ex}  1 & 5.88x10$^4$ & 2.53x10$^6$ \\
\hline
\rule[-1ex]{0pt}{3.5ex} 2 & 5.89x10$^4$ & 2.44x10$^6$  \\
\hline
\rule[-1ex]{0pt}{3.5ex}  3 & 6.29x10$^4$ & 1.80x10$^6$  \\
\hline
\rule[-1ex]{0pt}{3.5ex} 4 & 9.83x10$^4$ & 3.93x10$^6$  \\
\hline
\rule[-1ex]{0pt}{3.5ex}  5 & 8.88x10$^4$ & 1.47x10$^6$  \\
\hline
\rule[-1ex]{0pt}{3.5ex}  6 & 6.05x10$^4$ & 1.28x10$^6$  \\
\hline
\rule[-1ex]{0pt}{3.5ex}  7 & 5.06x10$^4$ & 1.07x10$^6$  \\
\hline
\end{tabular}
\end{center}
\end{table}

\subsection{Error Analysis on Counting}

This section focuses on the error analysis of the damaged area from the recently concluded hypervelocity test series carried out. 

\subsubsection{Entrance Surface Wound}

The experimentally derived value of $\kappa$ is the ratio of the diameter of the hole in the entrance membrane, $D_h$, to the diameter of the penetrator, $d_p$.

The variance in $\kappa$ is found by applying the law of error propagation to Equation \ref{Eq:kappa} which is written

\begin{equation}
\label{Eq:VarK1}
   {\sigma _{\kappa }^{2}={{\left( \frac{\partial \kappa }{\partial {{D}_{h}}} \right)}^{2}}\sigma _{{{D}_{h}}}^{2}+{{\left( \frac{\partial \kappa }{\partial {{d}_{p}}} \right)}^{2}}\sigma _{{{d}_{p}}}^{2}.}
\end{equation}

Substitution of the partial derivatives in Equation \ref{Eq:VarK1} gives

\begin{equation}
\label{Eq:VarK2}
  {\sigma _{\kappa }^{2}={{\left( \frac{1}{d_{p}^{2}} \right)}}\sigma _{{{D}_{h}}}^{2}+{{\left( \frac{D_{h}^{2}}{d_{p}^{4}} \right)}}\sigma _{\kappa }^{2}}. 
\end{equation}

Since the variances of $D_h$ and $d_p$ can be expressed as a fraction of their values by the ratios $g$ and $h$ respectively, we may write

\begin{equation}
\label{Eq:VarK3}
 \sigma _{\kappa }^{2}={{\left( \frac{1}{d_{p}^{2}} \right)}}\left( D_{h}^{2}{{g}^{2}} \right)+{{\left( \frac{D_{h}^{2}}{d_{p}^{4}} \right)}}\left( d_{p}^{2}{{h}} \right).  
\end{equation}

The value of $g$ is derived from the measurements of the diameter of the entrance wound. On the basis of repeated measurements of the entrance wounds on many different diameters, we estimate the that the precision on the measurement of $D_h$ is of order 0.03$D_h$ or 3$\%$. The value of $h$ is fractional error in $d_p$ and is the root-sum-square of the effects of non-sphericity, ~0.32$\%$ and average diameter error of 0.64$\%$.  These values combined give $h$ as 0.71$\%$.  So the estimated fractional error in $\kappa$ is approximately 3.1 $\%$

\subsubsection{Exit Surface Wound}

In this section we discuss our estimation of the uncertainty in determining the penetration area of the exit surface. This area is made up of many small holes counted and sorted into size bins as shown in Figure \ref{fig:6B_highdist} and referenced to the high or low density region. We can write the total area of penetration of the membranes,  $A_{exit, obs}$ as

\begin{equation}
\label{Eq:VarX1}
 {{A}_{exit,obs}}=4\left[ \left( \sum\limits_{i=1}^{w}{{{L}_{i}}\pi r_{i}^{2}} \right)+\left( 1-\gamma  \right)\left( \sum\limits_{i=1}^{w}{{{H}_{i}}\pi r_{i}^{2}} \right) \right].
\end{equation}
where the area of penetration in each quadrant is replaced by summations over a fixed number of bins, and the correction term $\gamma$ to avoid double counting. To formulate the variance in $A_{exit,obs}$, the law of error propagation is applied to Equation \eqref{Eq:VarX1} giving

\begin{equation}
\label{Eq:VarX2}
\sigma _{{{A}_{exit,obs}}}^{2}=\sum\limits_{i=1}^{w}{{{\left( \frac{\partial {{A}_{exit,obs}}}{\partial {{L}_{i}}} \right)}^{2}}\sigma _{{{L}_{i}}}^{2}}+\sum\limits_{i=1}^{w}{{{\left( \frac{\partial {{A}_{exit,obs}}}{\partial {{H}_{i}}} \right)}^{2}}\sigma _{{{H}_{i}}}^{2}}+{{\left( \frac{\partial {{A}_{exit,obs}}}{\partial \gamma } \right)}^{2}}\sigma _{\gamma }^{2}.
\end{equation}
Evaluation of the partial derivatives in Equation \ref{Eq:VarX2} gives
\begin{equation}
\label{Eq:VarX3}
 \sigma _{{{A}_{exit,obs}}}^{2}=\sum\limits_{i=1}^{w}{{{\left( 4\pi {{r}_{i}} \right)}^{2}}\sigma _{{{L}_{i}}}^{2}}+\sum\limits_{i=1}^{w}{{{\left( \left( 1-\gamma  \right)4\pi {{r}_{i}} \right)}^{2}}\sigma _{{{H}_{i}}}^{2}}+{{\left( 4{{A}_{q,H}} \right)}^{2}}\sigma _{\gamma}^{2}. 
\end{equation}
Reorganization of \ref{Eq:VarX3} gives
\begin{equation}
\label{Eq:VarX4}
 \sigma _{{{A}_{exit,obs}}}^{2}={{\left( 4{{A}_{q,H}} \right)}^{2}}\sigma _{\gamma }^{2}+4\pi \sum\limits_{i=1}^{w}{r_{i}^{2}\left( \sigma _{{{L}_{i}}}^{2}+\left( 1-\gamma  \right)\sigma _{{{H}_{i}}}^{2} \right)}.
\end{equation}
The leftmost term in Equation \ref{Eq:VarX4}, is due to variance in $\gamma$ and the terms under the summation signs are the expected variances in the counts of the various bins in the High and Low regions.  An experiment was performed to determine the variance in $\gamma$. The analyst divided up the area from a pristine image and then measured the $\gamma$ for each trial. The standard deviation in $\gamma$ was approximately 0.1$\gamma$ and substituted into Equation \ref{Eq:VarX4}. The variance in the counts in the bins $L_i$ and $H_i$ are determined by Poisson statistics and the variance is replaced by the mean and shown in Equation \ref{Eq:VarX5}. 

\begin{equation}
\label{Eq:VarX5}
 \sigma _{{{A}_{ent,obs}}}^{2}=0.01{{\left( 4{{A}_{q,H}} \right)}^{2}}+4\pi \sum\limits_{i=1}^{w}{r_{i}^{2}\left( {{L}_{i}}+\left( 1-\gamma  \right){{H}_{i}} \right)}.
\end{equation}
Since most of the area of the exit wound is found in bins with a large number of counts, the fractional variance is significantly less than 0.01, so the first term is dominant, and the standard uncertainty in the exit area is $\sim$10$\%$.

\subsection{Observed Impacts on the First Membrane and Determination of Parameter Beta}
\label{sect: first_membrane_tests}

In Section \ref{sect: mmod_entrance}, we reviewed the theoretical formulation for the hole diameter size as a function of the projectile properties, film properties, and impacting velocity (Equation \ref{Eq:hole_growth}). For this formula to be useful to extrapolate $\kappa$ to higher velocities, we need to determine the values for $\beta_1$ and $\beta_2$ in Equation \ref{Eq:B_coeff} for Kapton. In our experiments, all variables in Equation \ref{Eq:hole_growth} are held constant except for velocity, allowing us to determine the velocity dependence of $\beta$.

The hole diameter sizes measured on the entrance membranes give a corresponding ratio of $D_h/f \approx 20$. In this regime, Equation \ref{Eq:hole_growth} can be approximated as
\begin{equation}
   d_p^\prime \approx D_h^\prime \left( 1-exp(-D_h^\prime/\beta) \right).
\end{equation}
This can be written in terms of $\kappa$ as
\begin{equation}
   \frac{1}{\kappa} =  1-exp(-D_h^\prime/\beta).
\end{equation}
Solving for parameter $\beta$ then gives
\begin{equation}
\label{beta_approx}
   \beta=\frac{-D_h^\prime}{\ln(1-\frac{1}{\kappa})}.
\end{equation}

\begin{figure}
\begin{center}
\begin{tabular}{c}
\includegraphics[height=7cm]{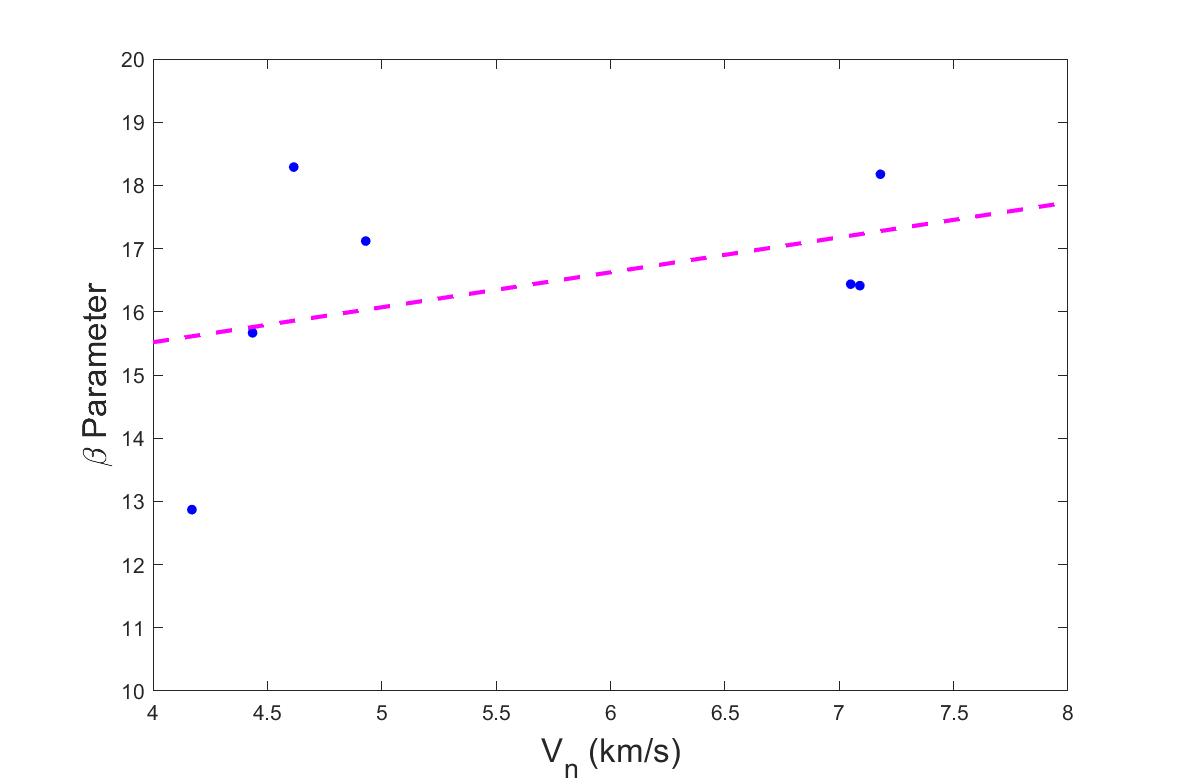}
\end{tabular}
\end{center}
\caption 
{\label{fig:B_fit}
Variation in B parameter with velocity. The dashed line is the line of best fit and is used to calculate the values for $\beta_1$ and $\beta_2$.}
\end{figure}

Since the first membrane displays damage from when the projectile was still intact, it is straight forward to calculate $\kappa$ for each shot using Equation \ref{Eq:kappa}.  The shots display a modest change in $\kappa$, ranging from 1.27 to 1.42  over this small velocity range, with an average value of $\kappa=$1.37. The calculated $\kappa$ and the observed hole size $D_h$ are  plugged into Equation \ref{beta_approx} to determine the corresponding $\beta$ parameter for each shot. Figure \ref{fig:B_fit} shows the variation in $\beta$ with velocity. The magenta dashed line in Figure \ref{fig:B_fit} is the line of best fit for the data points. We use this best fit line to describe how $\beta$ varies with velocity and find $\beta_1=13.3$ and $\beta_2=0.55$ in Equation \ref{Eq:B_coeff}. Thus, we can define $\beta$ for Kapton as $\beta=13.3 + 0.55V$, where the velocity is in units of $km/s$. We note here that this is a small sample size and more hypervelocity tests are needed to accurately determine $\beta_1$ and $\beta_2$. 

\begin{figure}
\begin{center}
\begin{tabular}{c}
\includegraphics[height=7cm]{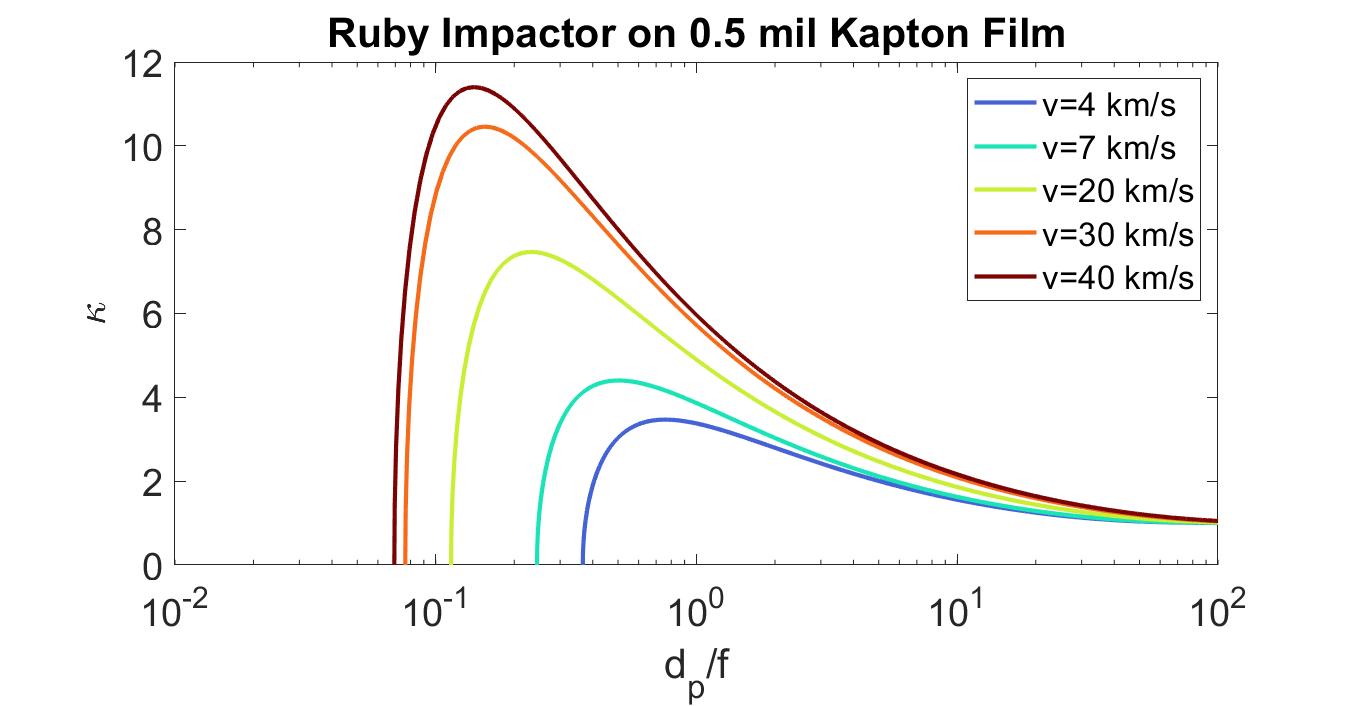}
\end{tabular}
\end{center}
\caption 
{ \label{fig:kappa_ruby}
Variation of $\kappa$ for a ruby impactor on a 0.5 mil Kapton film. $\kappa$ will vary with both the ratio of the projectile diameter ($d_p$) to the film thickness ($f$) and the velocity of the impactor.}
\end{figure}

Using the above definition for $\beta$, we can produce $\kappa$ curves for the experiment at different velocities using Equations \ref{Eq:hole_growth} and \ref{Eq:kappa}, shown in Figure \ref{fig:kappa_ruby}. We use values representative of Kapton and ruby, with inputs set to $f=12.7$ $\mu m$, $\sigma_t=8.79x10^7$ $Pa$, $\rho_t=1380$ $kg/m^3$, and $\rho_p=3950$ $kg/m^3$. These curves clearly demonstrate that impactors which are large relative to the membrane thickness exhibit minimal increases in $\kappa$ with an increase in velocity. The figure also illustrates how $\kappa$ rapidly increases as the projectile size becomes small relative to the membrane thickness, until it reaches the ballistic limit ($\kappa=0$). Our experimental set-up corresponds to an initial ratio of $d_p/f=15.75$  and range in velocity from ~4-7 $km/s$. However, after the micrometeoroid shatters, we expect its fragment distribution to span the full range of ratios less than this value, which correspond to maximum $\kappa$ values of ~3.5-4.5 over this velocity range. If future tests are performed at speeds more representative of micrometeoroids ($\sim$ 35 $km/s$), we expect the maximum $\kappa$ values for those experiments to be closer to 10.

\subsection{Observed Impacts on the Second Membrane and Implications for Fragmentation}
\label{sect: second_membrane_tests}

One of the goals of this paper is to empirically determine what the parameter values should be for micrometeoroid fragmentation. This requires examining the fragment distribution from our experiments to extract $\alpha_f$ and $c_{max}$ (which then determines $a_{f,max}$) which dictate this distribution. Although $a_{f,min}$ also influences the overall distribution, it does not affect the distribution as much as the other two parameters since that limit controls a small fraction of the total mass. Values acquired for $\alpha_f$ and $c_{max}$ can then be used to feed into a predictive model for the fragmentation.

The experiments performed directly measure the damage area caused by impacting fragments, not the fragment sizes themselves. Therefore, the number distribution of the fragments must be inferred. We utilize Equation \ref{Eq:hole_growth} to estimate the impacting fragment sizes that produced the observed holes on the second membranes. Since we do not know the individual velocities of each fragment, we make the simplifying assumption that all the fragments will have impacting speeds equal to that of the original impactor. The $d_p/f$ curves (similar to Figure \ref{fig:projectile_curve}) are produced for each shot's corresponding impact velocity to allow us to map the observed hole diameter $D_h$ to a corresponding fragment diameter $d_f$. 

A fragment array is created to bin fragment diameters ranging from 2 $\mu m$ (just below the ballistic limit for Kapton at these speeds) to 200 $\mu m$ (the original particle diameter) with a bin spacing of 4 $\mu m$. To arrive at an inferred number distribution, we calculate the $D_h/f$ ratio for each hole observed to determine the corresponding fragment diameter $d_f$ and add a number count to the bin encompassing the matched fragment size. A check is performed to ensure that all observed holes are assigned to a bin. Since the hole measurements are counted for a single quadrant, the number counts are multiplied by four to estimate the total number of fragments penetrating the membrane. We use a correction similar to Equation \ref{Eq:count_correction} to subtract the high density region contribution to the low density counts to avoid double counting holes.

\begin{figure}
\begin{center}
\begin{tabular}{c}
\includegraphics[height=10cm]{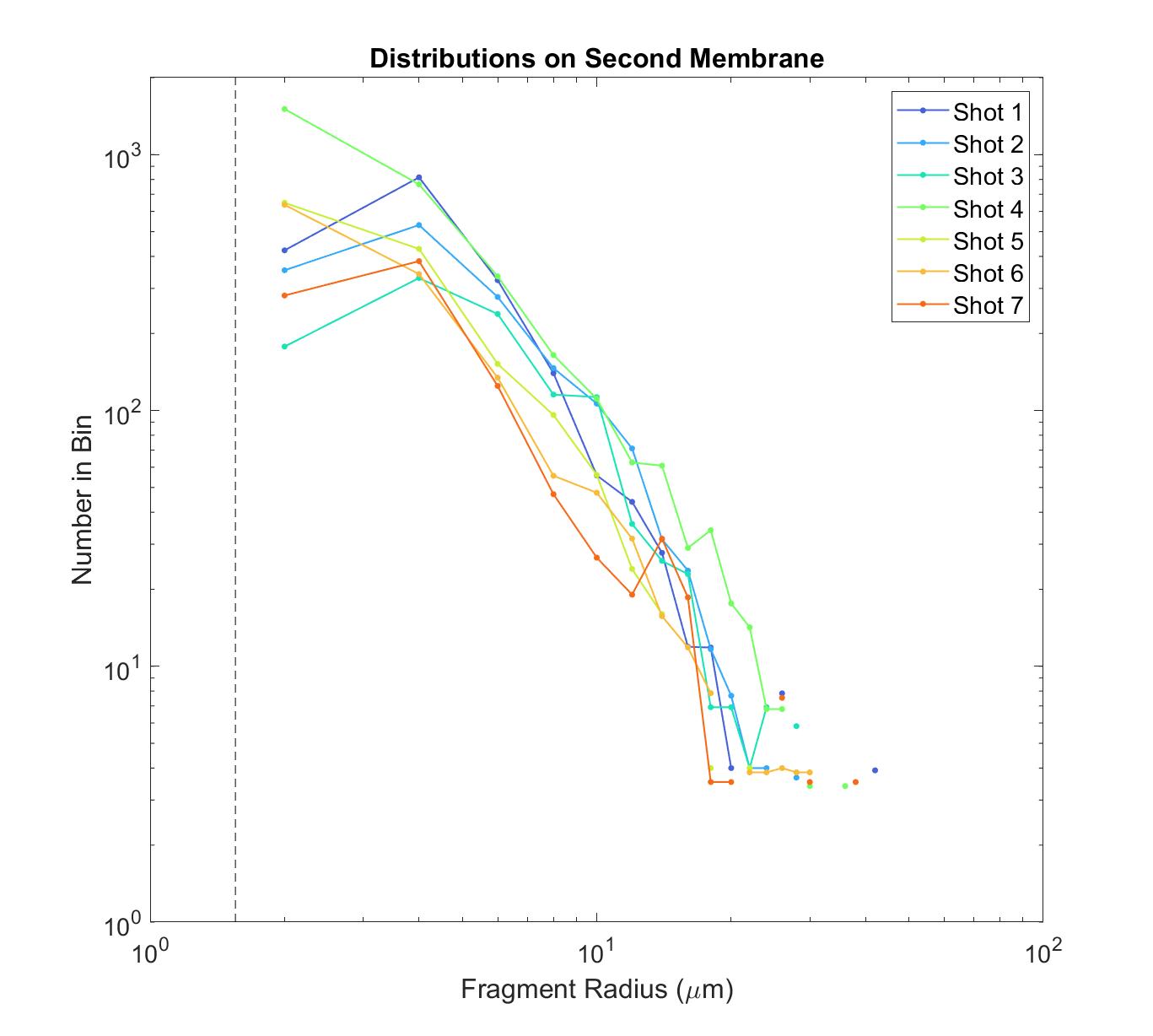}
\end{tabular}
\end{center}
\caption 
{ \label{fig:shot_distributions}
Fragment size is inferred from each damage hole observed. The fragments are then binned and counted. The dashed line indicates the ballistic limit for an impacting velocity of 7 $km/s$. Midpoints of the bins are used to plot the data.} 
\end{figure}

Figure \ref{fig:shot_distributions} shows the resulting number distributions from each impact test as a function of fragment radius. All the shots exhibit similar behavior for the fragmentation and follow a power law distribution as expected. The dashed line in Figure \ref{fig:shot_distributions} is the ballistic limit calculated using Equation \ref{Eq:ballistic_limit} for a reference velocity of 7 $km/s$. Note that there is a sharp drop-off in counts for the smallest fragment bins. This could be due to the fragments having a lower impact velocity than the original impactor (and thus $d_{p,bal}$ would be larger), particles not penetrating the membrane efficiently, difficulty observing the smallest holes, or all of the above.

\begin{figure}
\begin{center}
\begin{tabular}{c}
\includegraphics[height=6cm]{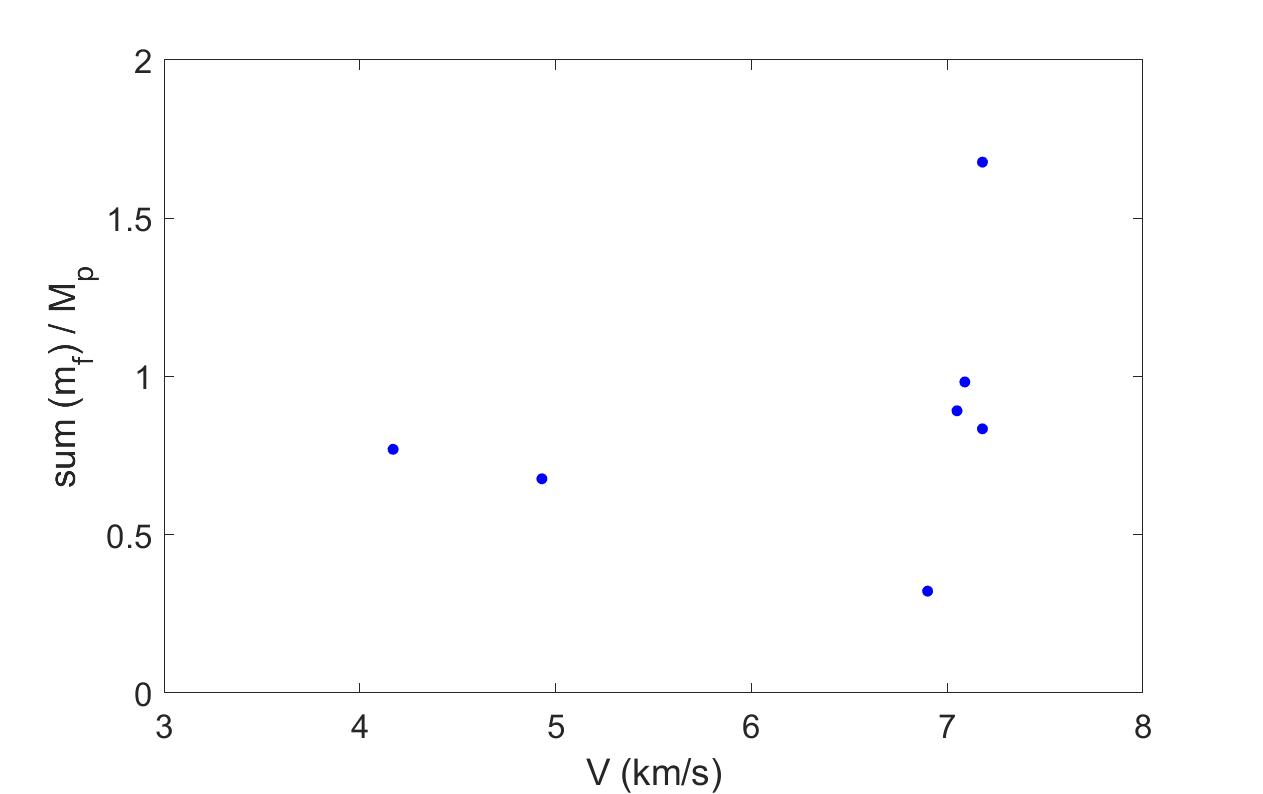}
\end{tabular}
\end{center}
\caption 
{ \label{fig:mass_check}
All the inferred fragment masses from Figure \ref{fig:shot_distributions} are summed to determine the total fragment mass. The ratio of the total fragment mass to the original projectile mass ($M_p$) is shown. These values should always be less than or equal to the projectile mass. } 
\end{figure}

As a sanity check to make sure our results make physical sense, we sum over all the fragment masses for each inferred number distribution and compare this value to the original impactor mass. This sum should always be less than or equal to the original impactor mass. We approximate the fragments as spheres and use the relation $m=\frac{4\pi}{3}\rho a_f^3$ to calculate the mass of each fragment. Figure \ref{fig:mass_check} displays this sanity check for each test shot. Almost all the shots are consistent with the conservation of mass, and have similar values for $(\sum m_f)/M_p$ with a median value of $\sim$0.8. Shots 4 and 5 appear to be anomalies, with mass fractions of about 1.6 and 0.31, respectively. Note that these are the two experiments where an incidence angle on the first membrane was imposed, which may be affecting the inferred fragment size. There are also other possible explanations for this error. The first is that we approximate all the projectile fragments as spheres. The shapes of the shards after every test completion were not examined, so it is hard to say how much the fragments deviate from a spherical shape. Future tests would be needed to determine their elongation. Other error contributions could be due to the assumptions in $\kappa$ and the count extrapolation from the quadrant method. Assuming Shots 1, 2, 3, 6, and 7 are good representations of the true fragment mass, Figure \ref{fig:mass_check} suggests about 17$\%$ of the mass is lost after the first membrane due to the inability of the smallest fragments to penetrate the second membrane at these speeds.

\begin{figure}
\begin{center}
\begin{tabular}{c}
\includegraphics[width=16cm]{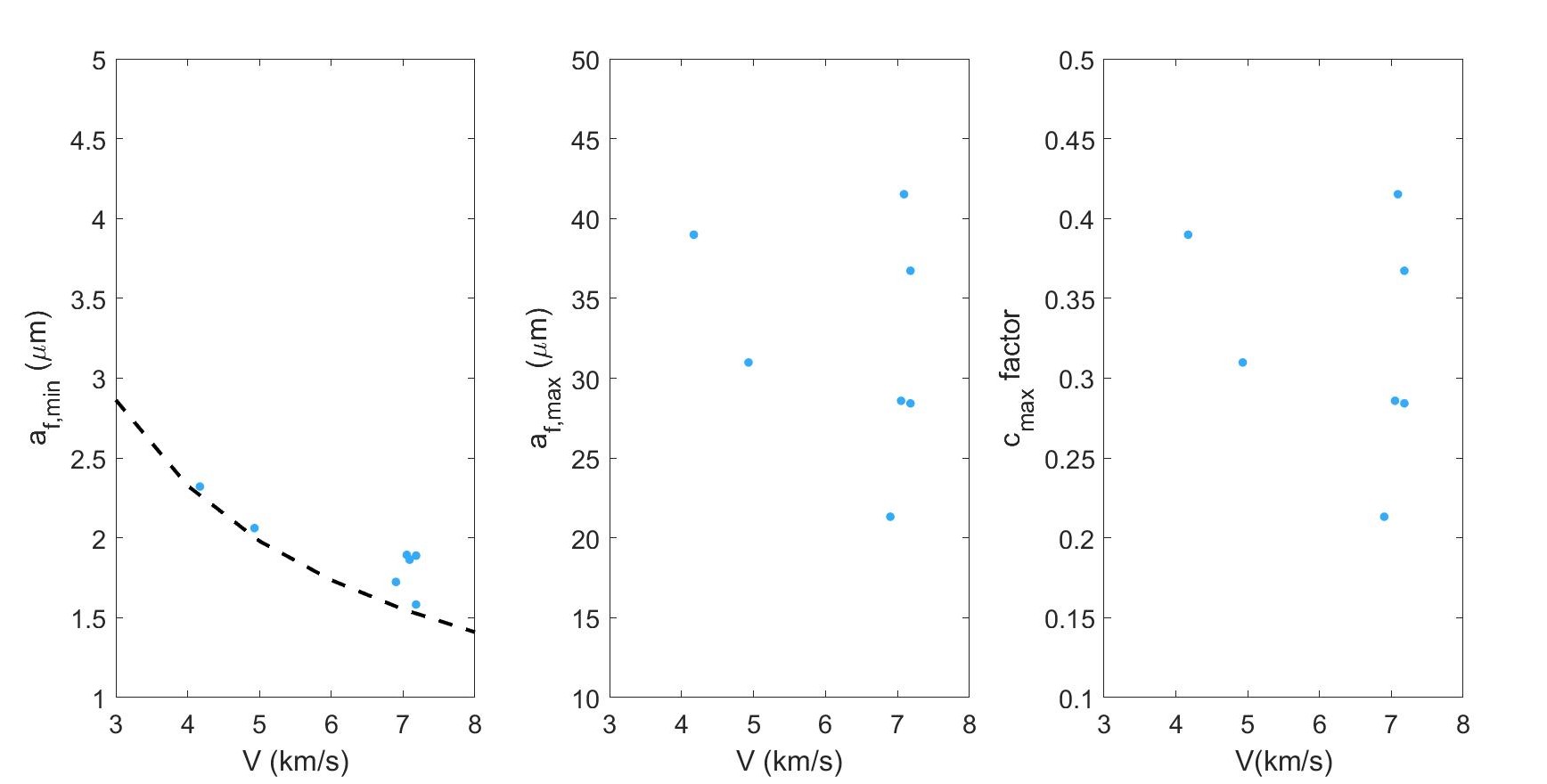}
\end{tabular}
\end{center}
\caption 
{ \label{fig:frag_panels}
Left panel: Minimum fragment size observed on each shot is near the ballistic limit (dashed line). Middle panel: Maximum fragment size observed on each shot. Right panel: Scaling factor derived based on maximum fragment size.} 
\end{figure}

In addition to the number distribution, we also want to determine the bounding limits of the fragmentation. The smallest and largest holes observed for each shot are identified and a corresponding fragment size is calculated. The left panel of Figure \ref{fig:frag_panels} shows that the smallest holes observed correspond to fragment sizes that are at or just above the ballistic limit (dashed line). This implies that the true value for $a_{f,min}$ of the fragment distribution is likely smaller than this, but cannot be observed since these fragments are not able to penetrate the membrane. The middle panel of Figure \ref{fig:frag_panels} shows that the test shots exhibit similar sizes for their observed $a_{f,max}$. Since the particle size is held constant throughout our experiments, this also gives rise to consistency in the calculated $c_{max}$ value. We find an average value of $c_{max}=$0.32, with the overall range encompassing the $c_{max}=$0.22 (Jones et al., 1996 \cite{Jones}) and $c_{max}=$0.27 (Hirashita and Kobayashi, 2013 \cite{hirashita}) values assumed in the literature.

\begin{figure}
\begin{center}
\begin{tabular}{c}
\includegraphics[height=10cm]{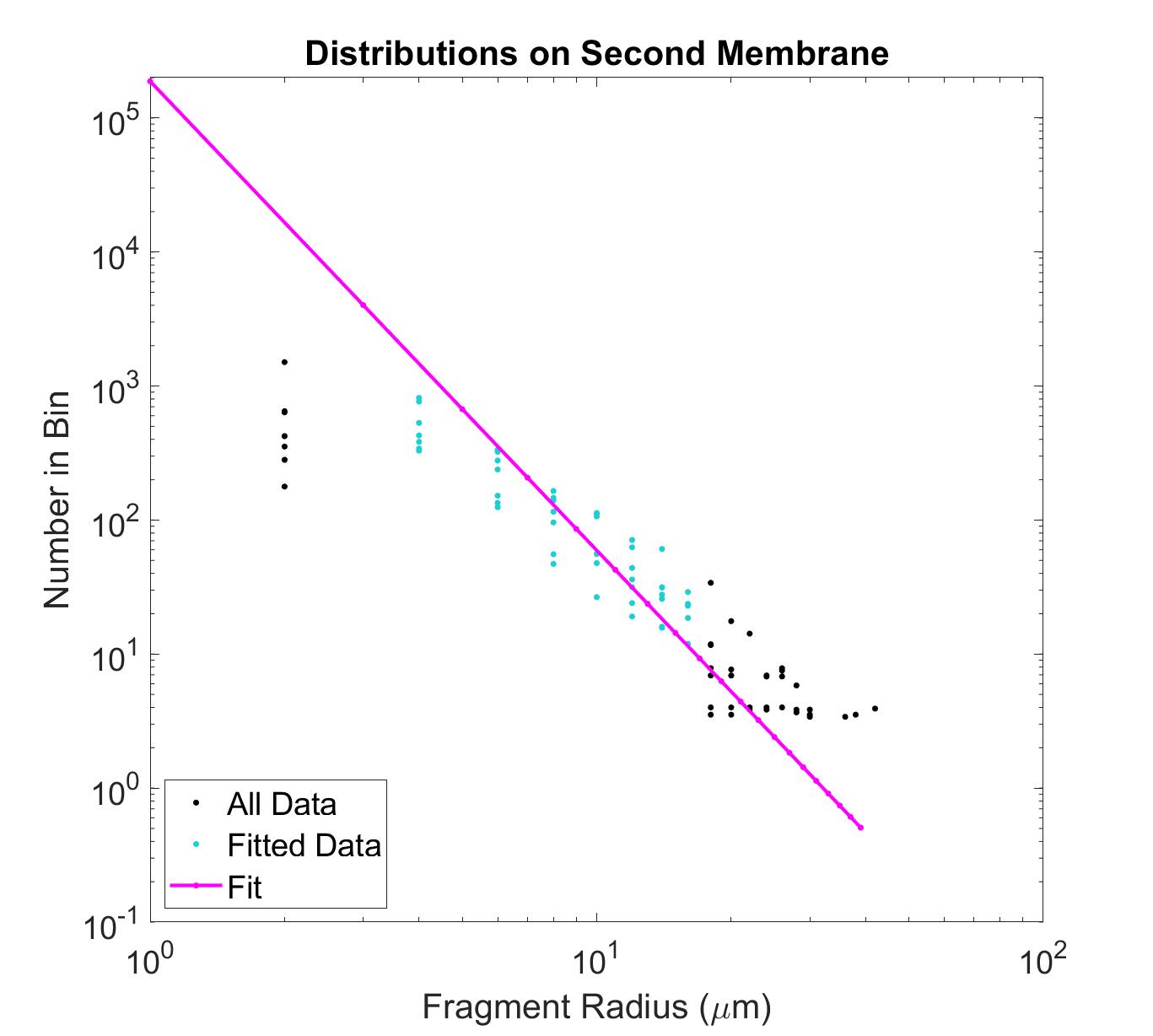}
\end{tabular}
\end{center}
\caption 
{ \label{fig:model_app}
The observed number distributions from the hypervelocity impact tests are best described by a fragmentation model with $\alpha_f=4$ and $c_{max}=0.24$. } 
\end{figure}

To establish a model that's representative of particle fragmentation incident on a polyimide film, we need to determine which combination of $\alpha_f$ and $c_{max}$ best replicates the observed fragmentation. The fragmentation is modeled using Equations \ref{Eq:nm_distribution}, \ref{Eq:a_max}, and \ref{Eq:C_f} to ensure conservation of mass. We use as inputs the original impactor mass, radius, and $\rho$ as defined by the experimental set-up (see Table \ref{tab:experiment_setup}). We set $a_{f,min}=1$ $\mu m$, just below the ballistic limit since the true value could not be determined. The $\alpha_f$ and $c_{max}$ parameters are varied to determine which combination produces the minimum residuals. The $\alpha_f$ parameter is varied from 3.0 to 5.5 with 0.1 increments and $c_{max}$ is varied from 0.20 to 0.40 with 0.01 increments. Since the smallest  and largest fragment bins have issues with their counting statistics, we limit the fit to fragments with radii between 3-18 $\mu m$. We find that the data is best fit by a fragmentation model with parameters $\alpha_f=4$, $c_{max}=0.24$, and $a_{min}=$1 $\mu m$. Figure \ref{fig:model_app} shows how the number distribution estimated from this model compares with the number distributions inferred from the observed impacts. Our estimate for $\alpha_f$ is slightly larger than the values used for interstellar grain fragmentation, which typically assume an $\alpha_f$ between 3.3-3.5. \cite{MRN} \cite{Jones}\cite{hirashita} The $c_{max}$ retrieved is intermediate between the literature values of 0.22 (assumed by Jones et al., 1996 \cite{Jones}) and 0.27 (assumed by Hirashita and Kobayashi, 2013 \cite{hirashita}).

\begin{figure}
\begin{center}
\begin{tabular}{c}
\includegraphics[height=7cm]{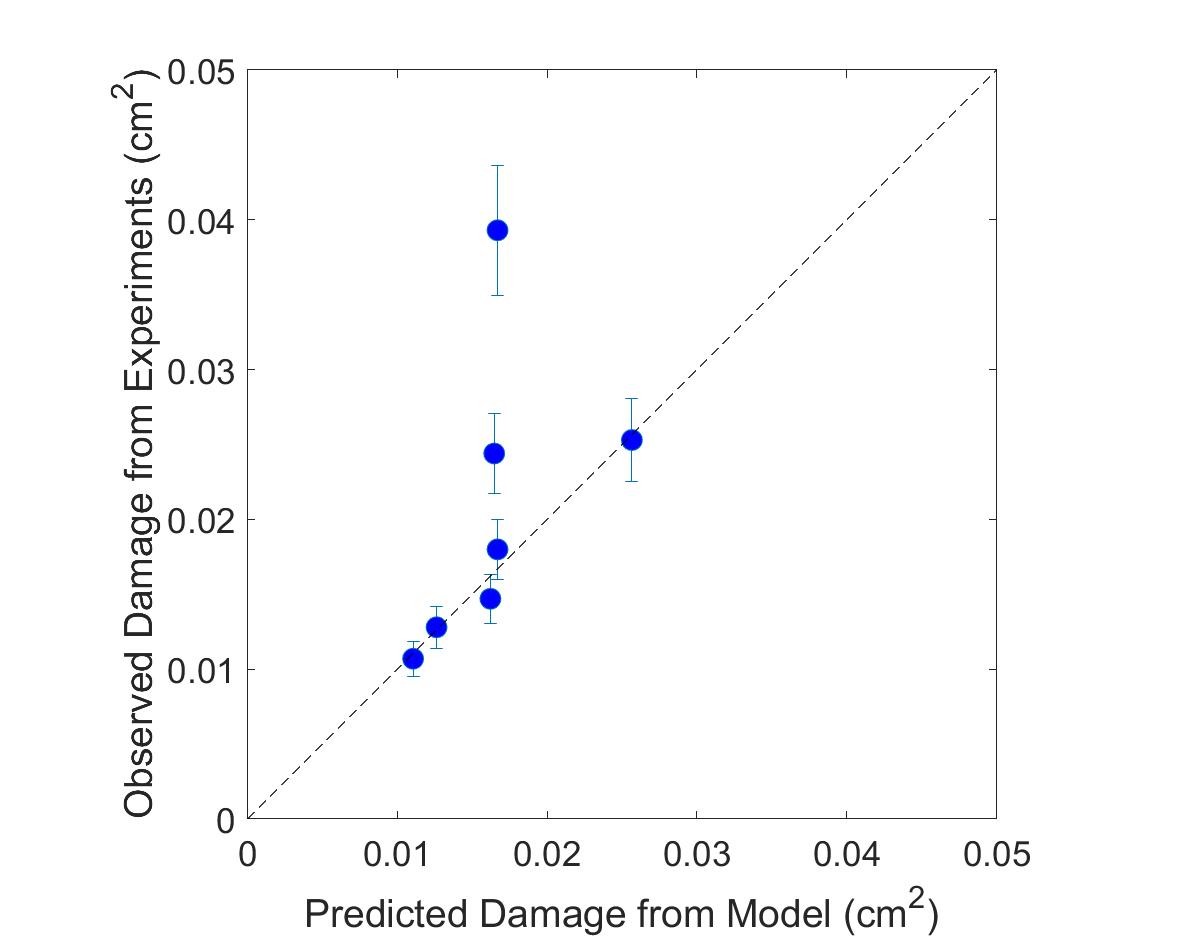}
\end{tabular}
\end{center}
\caption 
{ \label{fig:Damage_check}
Comparisons of the total damage to the second membrane predicted by our empirical model and that observed in our experiments. The dashed line represents the y=x line. } 
\end{figure}

As a sanity check, we use our empirical model to predict impact holes on the second membrane and compare with our observed results. We set $\alpha_f=4$, $c_{max}=0.24$, and $a_{f,min}=$1 $\mu m$ to recreate a number distribution for the shattering fragments impacting the second membrane. We use the velocities listed in Table \ref{tab:experiment_setup} to determine the appropriate $\kappa$ for each fragment (we assume the fragments have the same speed as the original impactor) and use Equation \ref{Eq:A_exit} to predict what the corresponding total damage to the second membrane should be. Figure \ref{fig:Damage_check} shows how well our model replicates the observed values from our experiments. For the most part, our model replicates the observed damage fairly well. The outlier data point near (0.017,0.039) is shot four, which we acknowledged earlier as an an anomalous data point, perhaps due to the effects of the inclined first membrane that are not captured by our model.

\subsection{Predictions for Future Tests with Multiple Membranes}
\label{sect: predictions}

One option to conserve inflatant and extend the mission lifetime is to construct a robust design. A natural reaction would be to consider thicker membranes. This is likely not the solution: to achieve the relevant stress in the film needed to make a good reflector, the pressure would need to be increased for a thicker membrane. In Section \ref{sect: gas_loss_deriv}, we will show that the gas loss through holes (Equation \ref{Eq41}) is proportional to the product of the hole area and the pressure. Thus, the cummulative hole area would have to fall off faster than linearly to extend the lifetime. Figure \ref{fig:kappa_ruby} displays that the film thickness would have to be significantly increased for the $\kappa$ (which will control the overall damage) to be significantly reduced for the smallest impacting particles which enact the most damage relative to their size. This will become more clear in Section \ref{sect: mitigation} where we consider different membrane thicknesses.

An alternative to thicker membranes is the option of additional membrane layers to shield the central structure. These layers are added to cause successive fragmentation and reduce penetration of the lenticular.  The effectiveness of such a shield will need to be validated with additional hypervelocity tests. In anticipation of such tests, we use the empirical model from the previous section to predict the damage the fragments will cause on each subsequent layer. This will allow us to test our fragmentation model in future laboratory experiments. While we keep the velocity constant to show the effects of increasing particle size for clarity, the inputs into the predictive model can be adjusted to forecast the damage for any experimental design. 

To estimate the damage of a ruby projectile on N number of layers, we apply our model using $\alpha_f$=4, $c_{max}$=0.24, $a_{f,min}=$1 $\mu m$, impacting velocity $V=$7 $km/s$, impacting angle $\theta=$0 $deg$, and a variable $\kappa$ dependent on projectile diameter to film thickness  ($V=$7 $km/s$ $\kappa$ curve in Figure \ref{fig:kappa_ruby}). We set each layer to the same thickness of 0.5 $mil$ (12.7 $\mu m$) and use values representative of Kapton for their composition. In this analysis, we will consider a set-up of four consecutive layers.

\begin{figure}
\begin{center}
\begin{tabular}{c}
\includegraphics[height=8cm]{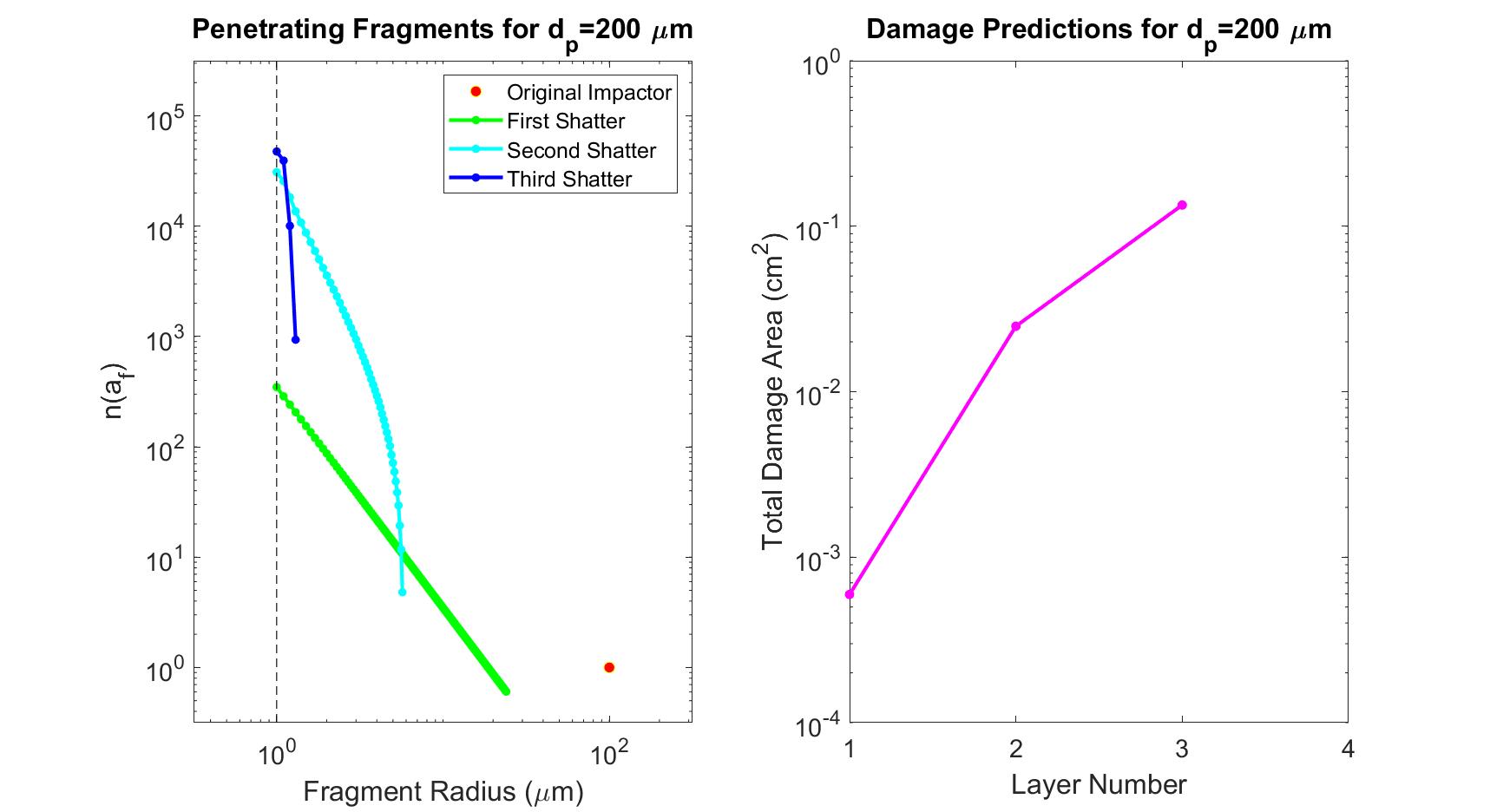}
\end{tabular}
\end{center}
\caption 
{ \label{fig:layers_example}
Predictions for a 200 $\mu m$ diameter ruby impacting four layers of 0.5 $mil$ Kapton at 7 $km/s$. Left panel: Number distributions for fragments capable of penetrating the membrane after each subsequent shatter. The dashed line indicates the ballistic limit for the membrane. Right: Total damage predicted on each subsequent membrane layer.} 
\end{figure}

We estimate the damage for four initial projectile diameter sizes: 100 $\mu m$ (2.07x10$^{-6}$ $g$), 200 $\mu m$ (1.65x10$^{-5}$ $g$), 1000 $\mu m$ (2.07x10$^{-3}$ $g$), and 2000 $\mu m$ (1.65x10$^{-2}$ $g$). We utilize Equations \ref{Eq:nm_N_layers} and \ref{Eq: A_damage_N_layers} to calculate the expected fragmentation and corresponding damage area upon each membrane. Figure \ref{fig:layers_example} shows an example for a 200 $\mu m$ incoming particle, the same as that used in our laboratory tests from Section \ref{sect: hypervelocity_tests}. The left panel shows the remaining distribution of the fragments after each layer impact. The number of fragments sharply increases as the micrometeoroid is pulverized into smaller grains. However, as these fragments become smaller and smaller, they do not possess enough momentum to exceed the threshold set by the ballistic limit (dashed line in Figure \ref{fig:layers_example}), resulting in mass loss after each interface. The right panel of Figure \ref{fig:layers_example} shows the total damage incurred by each layer. As the fragments become more numerous and smaller (larger $\kappa$), there are more punctures on each consecutive membrane, corresponding to an increase in damage incurred. However, there is a turning point at which enough mass is lost that the damage begins to decrease. For this example, the third layer is the most perforated, while the particles thereafter are too small to penetrate the fourth layer.

\begin{figure}
\begin{center}
\begin{tabular}{c}
\includegraphics[height=8cm]{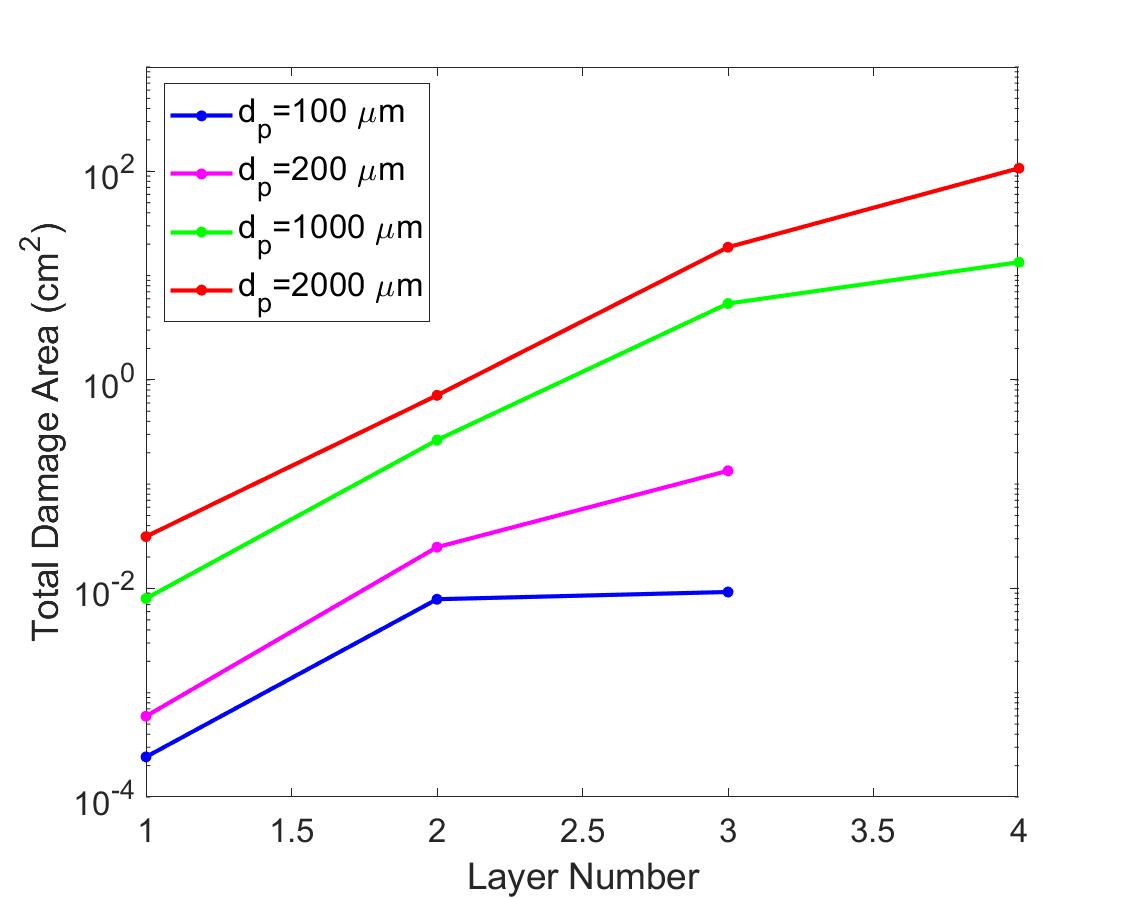}
\end{tabular}
\end{center}
\caption 
{ \label{fig:layers_all}
The total damage area on each membrane layer (0.5 mil Kapton) is predicted for four ruby particle sizes. } 
\end{figure}

Figure \ref{fig:layers_all} shows the results for damage caused by a ruby projectile with a diameter of 100, 200, 1000, or 2000 $\mu m$. The model predicts that there will be no fragments capable of penetrating the fourth membrane for an initial projectile sizes of 100 and 200 $\mu m$. However, the larger particles at 1000 and 2000 $\mu m$ penetrate all four layers.

Note that in this analysis, we assume the fragments maintain the 7 $km/s$ speed after each interface to determine the ballistic limit and corresponding $\kappa$. Since the fragments will lose energy upon each impact, the predicted damage areas are a bounding worst case scenario. In addition, we have set the minimum fragment radius to be 1 $\mu m$. The actual minimum fragment radius may be smaller than this, and therefore mass could be lost at each interface quicker than that predicted here. More experiments are needed to establish how efficiently micrometeoroid grains are removed after each obstruction.

\section{Micrometeoroid Mitigation}
\label{sect: mitigation}

In Section \ref{sect: gas_loss_deriv} we will find that the total gas mass required to accommodate the inflated lenticular is proportional to the mission lifetime squared. This requirement is heavily dependent on the rate at which the lenticular accumulates holes in its surface due to micrometeoroid impacts. Therefore, if we can reduce the hole accumulation rate, we can extend the lifetime of the inflatant substantially. In this section, we investigate how many layers would be required to create a micrometeoroid shield capable of mitigating incoming micrometeoroids at a location near 1 AU. In addition, the same dynamics will occur for structures with multiple membranes, such as sunshields, allowing the analysis to serve multiple functions.

\begin{figure}
\begin{center}
\begin{tabular}{c}
\includegraphics[height=8cm]{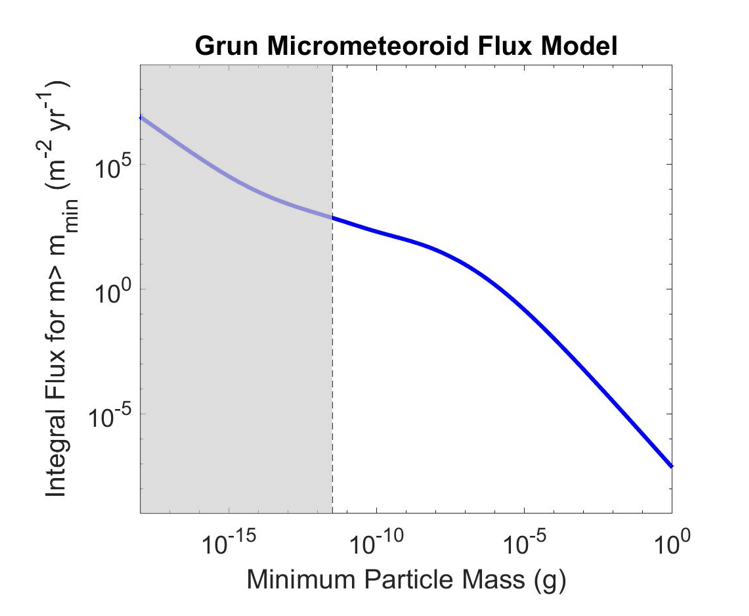}
\end{tabular}
\end{center}
\caption 
{ Grun et al. (1985) model for the micrometeoroid flux near 1 AU. Micrometeoroids less than 10$^{-12}$ $g$ (shaded gray region) with impact speeds of $\leq$35 $km/s$ will not penetrate a 0.5 $mil$ (or thicker) Kapton film. }
\label{fig:Grun_flux}
\end{figure}

To assess which micrometeoroids will be the most problematic, we employ the sky-averaged micrometeoroid flux model by Grun et al. (1985) \cite{GRUN} to describe the micrometeoroid environment at 1 AU. This model is valid for micrometeoroid masses 10$^{-18}<m_p<$1 $g$ and characterizes the integral micrometeoroid flux as a function of minimum threshold mass $m_p$ as
\begin{equation}
       \label{Eq: Grun_model}
    \Phi(m_p)=3.15576x10^7 \left(F_1(m_p) + F_2(m_p)+ F_3(m_p) \right)
\end{equation}
where
\begin{align}
    F_1(m_p)=(2.2x10^3 m_p^{0.306} + 15)^{-4.38} ,\\
    F_2(m_p)=1.3x10^{-9}(m_p+10^{11} m_p^2+10^{27} m_p^4)^{-0.36} ,\\
    F_3(m_p)=1.3x10^{-16}(m_p+10^6 m_p^2)^{-0.85}.
\end{align}
Figure \ref{fig:Grun_flux} shows the micrometeoroid flux near 1 AU. About 90$\%$ of micrometeoroids near 1 AU will impact with speeds of 35 $km/s$ or less (Thorpe et al., 2016 \cite{Thorpe2016}). At a speed of 35 $km/s$, a large portion of the micrometeoroid flux ($m_p<$10$^{-12}$, gray region in Figure \ref{fig:Grun_flux}) will not have enough momentum to penetrate a 0.5 $mil$ Kapton membrane. Hence, a significant portion of the micrometeoroid population is mitigated due to their relative speeds alone. The remaining population presents a hazard to the inflatable lenticular. There is $\sim$10 orders of magnitude difference in flux between particles at the ballistic limit and the maximum 1 $g$ size considered, with a rapid drop in flux for micrometeoroids larger than 10$^{-8}$ $g$.

Although in Section \ref{sect: predictions} we found that the total damage area on each successive membrane quickly scales with the original mass of the incoming micrometeoroid (Figure \ref{fig:layers_all}), impacts due to large-sized micrometeoroids occur much less frequently. For example, Figure \ref{fig:Grun_flux} indicates that micrometeoroids greater than 10$^{-5}$ $g$ occur at a rate of less than 1 per year per $m^2$. Hence, there is a trade-off between size and frequency which must be considered to determine how quickly a micrometeoroid shield will incur damage. 

\begin{figure}
\begin{center}
\begin{tabular}{c}
\includegraphics[height=7cm]{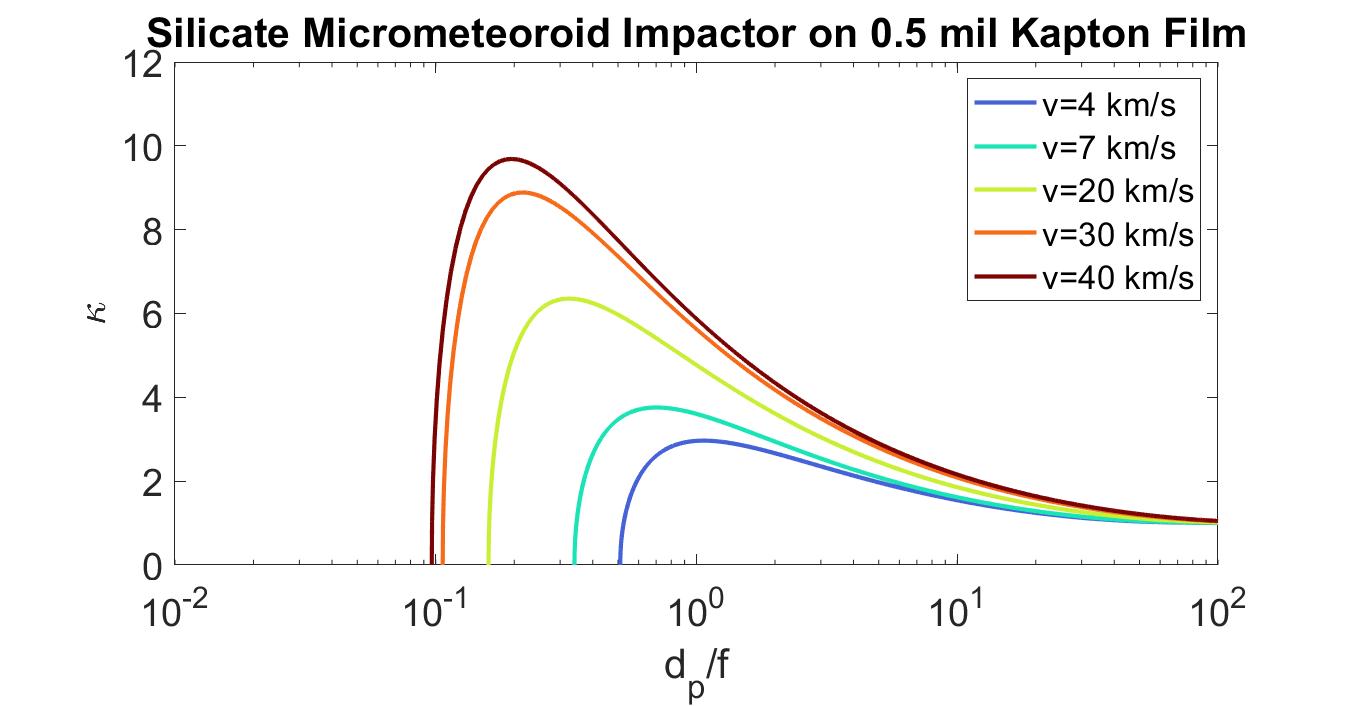}
\end{tabular}
\end{center}
\caption 
{ \label{fig:kappa_silicate}
Variation of $\kappa$ for a silicate micrometeoroid impactor on a 0.5 mil Kapton film. $\kappa$ will vary with both the ratio of the  to the film thickness ($f$) and the velocity of the impactor.}
\end{figure}

We utilize the information gathered from our laboratory tests and apply it to the real-world space environment. The interaction is modeled for the case of micrometeoroids impacting a lenticular membrane protected by sacrificial shield layers that are all composed of Kapton sheets, each with the same thickness. The parameter values for the fragmentation obtained in Section \ref{sect: second_membrane_tests} are used to extrapolate results for velocities reflective of micrometeoroids. We assume that a silicate micrometeoroid ($\rho$=2.5 $g/cm^3$ \cite{GRUN})) fragments in the same manner as that observed for the ruby test particles, with an $\alpha_f$=4, $c_{max}$=0.24, and an $a_{f,min}$=1 $\mu  m$. We implement the $\beta$ parameter observed for Kapton in Section \ref{sect: first_membrane_tests} and set $\beta_1=13.3$ and $\beta_2=0.55$.

\begin{figure}
\begin{center}
\begin{tabular}{c}
\includegraphics[height=8cm]{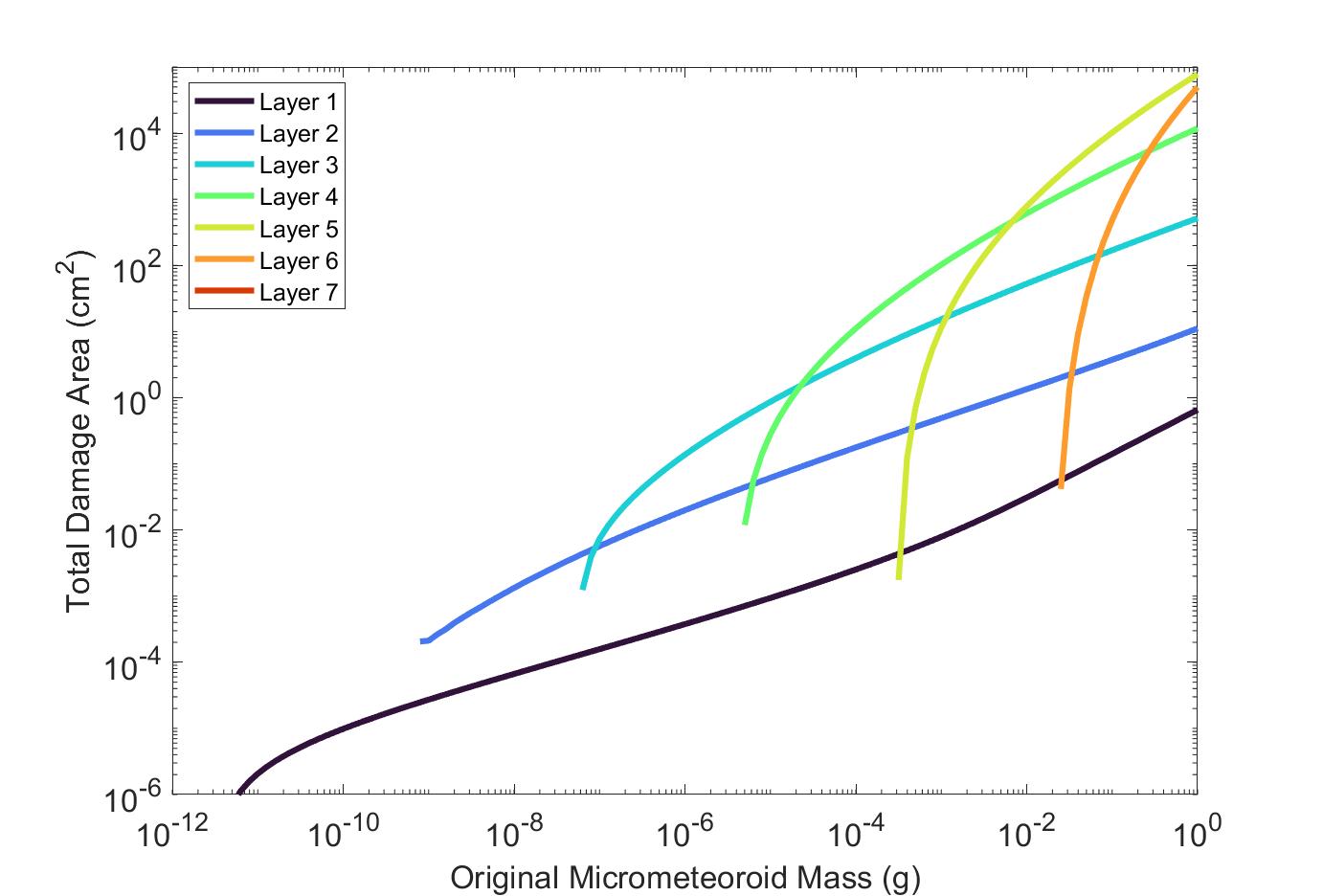}
\end{tabular}
\end{center}
\caption 
{ \label{fig:damagepermembrane}
Total damage due to fragments on each consecutive (0.5 $mil$) membrane based on the mass of the original micrometeoroid impactor. Mass is lost after each interface due to the inability of the smallest fragments to penetrate the membrane.} 
\end{figure}

The total damage is estimated for three cases of  membrane thickness: 0.5 $mil$ (12.7 $\mu m$), 1 $mil$ (25.4 $\mu m$), and 2 $mil$ (50.8 $\mu m$). Analysis is performed for an incidence angle of $\theta$=0, a velocity reflective of micrometeoroids near 1 AU of $V$=35 $km/s$ \cite{Thorpe2016}, and for particle masses of 10$^{-12}$ to 1 $g$. The lower bound of this mass limit corresponds to the ballistic limit for the 0.5 $mil$ thickness case (the other cases require a minimum mass larger than this to penetrate). We use Equations \ref{Eq:hole_growth} and \ref{Eq:kappa} to determine the $\kappa$ values for all $d_p/f$ ratios, shown in Figure \ref{fig:kappa_silicate}. Fragments are assumed to maintain the same speed as the original impactor. Equation \ref{Eq: A_damage_N_layers} is then used to determine the resulting damage on each wall. Figure \ref{fig:damagepermembrane} shows the damage area caused on successive 0.5 $mil$ layers for an incoming micrometeoroid of mass $m_p$. The increasing drop-off in the damage distribution depicts the survivability of the fragments and demonstrates the minimum micrometeoroid mass needed to reach a layer $N$. For example, micrometeoroids $<$10$^{-8}$ $g$ are successfully eliminated after the second layer. The figure also shows the substantial damage the fragments from a very large ($\sim 1$ $g$) micrometeoroid can cause to each layer.

\begin{figure}
\begin{center}
\begin{tabular}{c}
\includegraphics[height=8cm]{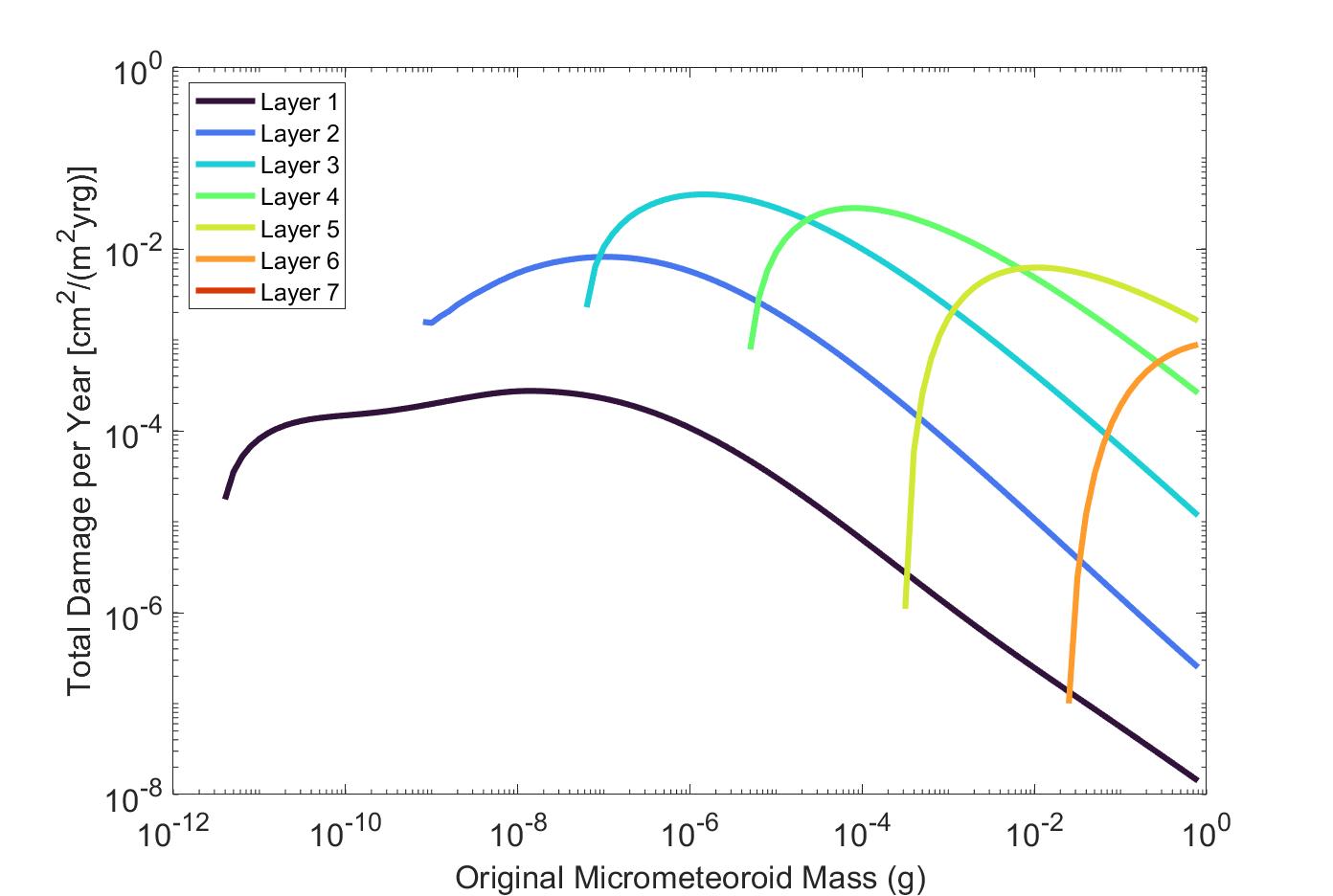}
\end{tabular}
\end{center}
\caption 
{ \label{fig:damage_rate}
The Grun flux model at the 1 AU location and the damage per micrometeoroid calculated in Figure \ref{fig:damagepermembrane} are used to estimate the damage rate on each 0.5 mil membrane for a given impacting micrometeoroid size. } 
\end{figure}

To determine the damage rate per year for a given mass, we use Equation \ref{Eq: Grun_model} to calculate the differential fluxes for each mass and multiply this by the damage per particle.  Figure \ref{fig:damage_rate} shows the damage that would be caused per year for a micrometeoroid at a given mass for the 0.5 $mil$ case. The plot indicates that micrometeoroids between the range $10^{-8}$ to $10^{-4}$ $g$ will be the most problematic since they have a balance between occurrence and damage caused.

\begin{figure}
\begin{center}
\begin{tabular}{c}
\includegraphics[height=7cm]{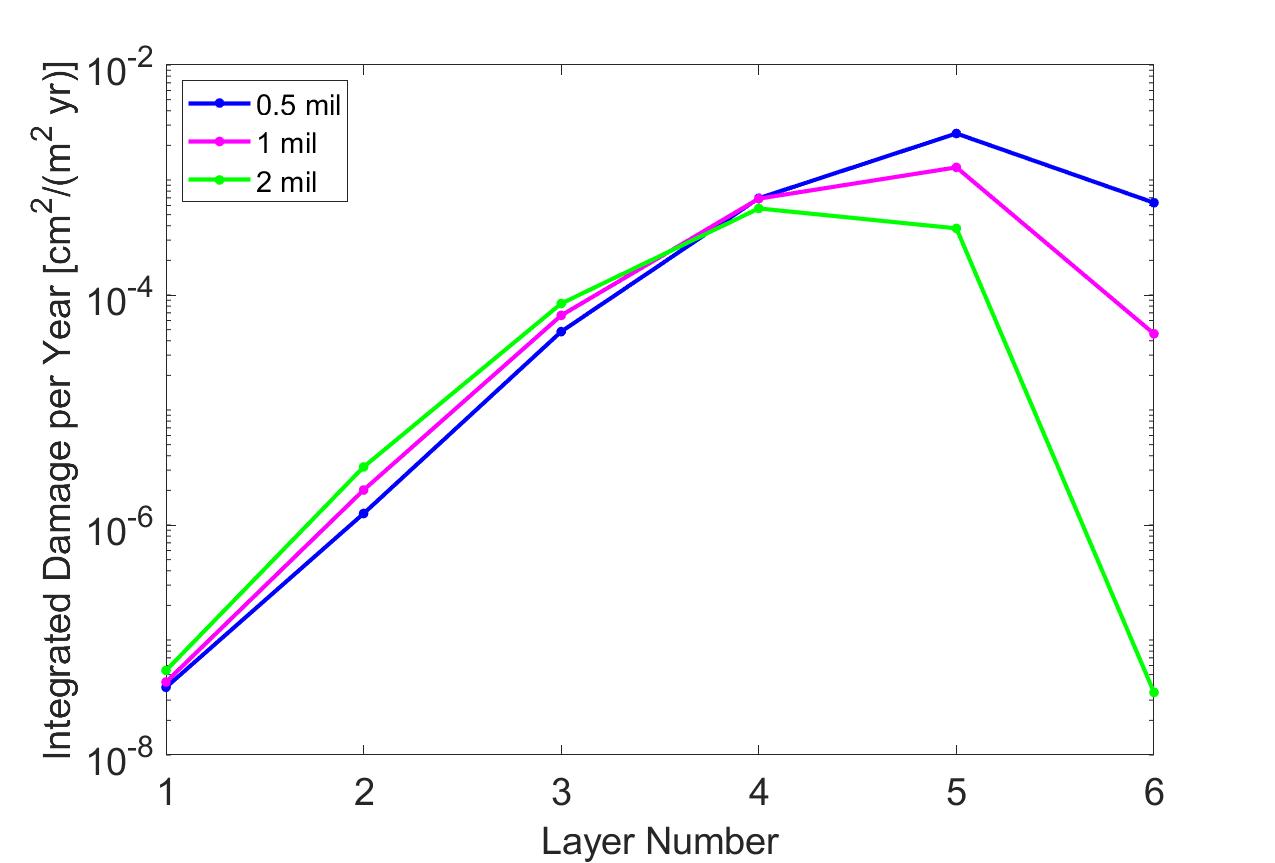}
\end{tabular}
\end{center}
\caption 
{ \label{fig:damage_integrated}
The damage rate per micrometeoroid mass in Figure \ref{fig:damage_rate} is integrated across all micrometeoroid masses to determine the total damage per year per $m^2$ . } 
\end{figure}

The damage rates per particle size shown in Figure \ref{fig:damage_rate} are integrated over to determine the total damage rate per year that would be experienced by each membrane. This corresponds to the integral term presented in Equation \ref{Eq42}. Figure \ref{fig:damage_integrated} shows the results for film thicknesses of 0.5, 1, and 2 $mil$. The fourth and fifth layers of the micrometeoroid shield sustain the most damage. The micrometeoroid fragments are successfully eliminated after the sixth layer for a film thickness of 2 $mil$ and after the seventh layer for film thicknesses of 0.5 and 1 $mil$. Therefore, to adequately protect against micrometeoroids will require several layers of Kapton sheets.

Note that Figure \ref{fig:damage_integrated} can also be used to determine the damage to an unshielded lenticular: Layers 1 and 2 can be thought of as the entrance and exit membranes, respectively. The authors would like to highlight the effect of a thicker membrane in the case of an unshielded lenticular: a thicker membrane has the opposite effect desired and actually increases the total damage areas $A_{ent}$ and $A_{exit}$ since $\kappa$ is amplified at these $d_p/f$ ratios for the incoming micrometeoroid distribution. The integral term in Equation \ref{Eq42} can be estimated using values from Figure \ref{fig:damage_integrated} for the 0.5 $mil$ case,  which results in a rate of 1.3x10$^{-6}$ $\frac{cm^2}{m^2 yr}$ near the 1 AU location. This rate can then be plugged into Equation \ref{Eq42} to determine the total hole area with time and subsequently estimate the gas required to complete the mission (covered in the next section).

\section{Mass Loss from Punctures and Permeation}
\label{sect: gas_loss_deriv}

The efficacy of the inflatable structure to operate as a primary reflector is rooted in the ability to maintain a static shape through constant pressure. This can be accomplished through a pressure maintenance system which will apply corrections to account for any changes in the number of gas molecules or  temperature. We expect drops in pressure to occur as the gas escapes through tiny punctures caused by micrometeoroids, which will require the pressure maintenance system to replenish the gas. To derive a function to describe the flow rate out of the lenticular, we need to determine which regime describes the flow. In addition, we also consider the possibility of gas escaping through permeation. We will use these descriptions as our basis for determining the total gas required to maintain the reflector over the course of the mission lifetime.

 At a typical operating temperature and pressure for an inflatable optic of 300 $K$ and 3.5 $Pa$, the mean free path for a gas will be on the order of 4 $mm$. Using a micrometeoroid speed of 35 $km/s$ and computing the corresponding $\kappa$ for an impact on 0.5 $mil$ Kapton (Figure \ref{fig:kappa_silicate}), we find that the mean free path is considerably larger than the hole diameters caused by micrometeoroids for particle sizes $\leq$10$^{-4}$ $g$. Micrometeoroids greater than this mass have a low occurrence rate; therefore, the gas flow caused by the majority of punctures can be described by effusion.

The effusion regime describes the flow rate $Q_{eff}$ out of a small hole area $A_H$ (defined in Equation \ref{Eq42}) as \cite{Liboff} 
\begin{equation}
    \label{Eq40}
    Q_{eff}=P A_H \sqrt{\frac{1}{2 \pi m_{molec} k_B T}}.
\end{equation}
where $P$ is the gas pressure, $m_{molec}$ is the mass per molecule, $k_B$ is the Boltzmann constant, and $T$ is the temperature of the gas. To get the rate of mass lost, $\Dot{m}=\frac{dm}{dt}$, $Q_{eff}$ is multiplied by  the mass per molecule, $m_{molec}$
\begin{equation}
    \label{Eq41}
    \Dot{m}=Q_{eff}m_{molec}=-PA_H\sqrt{\frac{M}{2 \pi N_A k_B T}}
\end{equation}
where $M$ is the molar mass, $N_A$ is Avogadro’s number, and $m_{molec} =\frac{M}{N_A}$. Note the dependence of Equation \ref{Eq41} on temperature: in the format we have written the equation, the mass loss rate is inversely proportional to the temperature. This means gas will escape slower for hotter systems, and faster for colder systems, which initially seems counterintuitive. However, if the system maintains constant pressure and volume, which will be imposed for an inflatable system, then temperature increases are accompanied by a decrease in the number of molecules in the balloon, which lead to less collisions, and vice versa.

As part of the mission design, all the terms in the Equation \ref{Eq41} will remain constant, with the exception of the hole area and gas temperature. The gas temperature will have a limited range due to the sun illumination angles allowed by the mission design, while the hole area $A_H$ will continually increase with time. Recall that the micrometeoroid hole accumulation (Equation \ref{Eq42}) is proportional to the diameter of the reflector. Consequently, larger diameter designs will have significantly shorter lifetimes for the same allocated gas budget compared to their smaller diameter counterparts since they lose gas more efficiently. This results in a trade-off between aperture size and mission lifetime. 

Now that we have defined the function for gas escape and the terms it contains, we can quantify the total gas needed over a mission lifetime, $Z$. Substituting the hole accumulation rate, Equation \ref{Eq42}, into Equation \ref{Eq41} and substituting in the gas constant, $R=N_A k_B$, we can rewrite the time rate of mass loss as 
\begin{equation}
    \label{Eq44}
    \frac{dm}{dt}=-A_{seams}P\sqrt{\frac{M}{2 \pi R T}}- A_{R}P\sqrt{\frac{M}{2 \pi R T}} t \int_{m_{p,min}}^{m_{p,max}} \Phi(m_p) A_{impact} dm_p.
\end{equation}
As complicated as Equation \ref{Eq44} looks it is very simple, namely 
\begin{equation}
    \label{Eq45}
    \frac{dm}{dt}=-(yt + j)
\end{equation}
where $y$ is given by 
\begin{equation}
    \label{Eq46}
    y=A_{R}P\sqrt{\frac{M}{2 \pi R T}}  \int_{m_{p,min}}^{m_{p,max}} \Phi(m_p) A_{impact} dm_p
\end{equation}
And $j$ is
\begin{equation}
    \label{Eq47}
    j=A_{seams}P\sqrt{\frac{M}{2 \pi R T}}
\end{equation}
with a bounding lower value used for $T$. In Equations \ref{Eq46} and \ref{Eq47}, all of the terms are set by the mission design and are therefore known, except for the integral term in Equation \ref{Eq46}, which can now be approximated using our predictive model.  The authors would like to highlight the implications of the dependence of $dm/dt$ on $yt$ in Equation \ref{Eq45}: since micrometeoroid holes increase with time, observations should be maximized early in the mission when $dm/dt$ is small, especially for objects which require long observation times. As the mission progresses, the same target integration time will be become more and more expensive to perform. 

The total gas mass needed for the mission is the mass needed to replace the total mass lost, given by 
\begin{equation}
    \label{Eq48}
    m_{G}=\int_{0}^{Z} (yt + j) dt
\end{equation}
which trivially integrates to 
\begin{equation}
    \label{Eq49}
    m_{G}=\frac{y}{2}Z^2 + jZ.
\end{equation}
As Equation \ref{Eq49} clearly shows, the inflatant mass required is proportional to the square of the mission length, $Z^2$. The expression we have found is valid over long timescales. However, on short timescales ($\sim$1 hour), there will be small deviations in the pressure (but still within the pressure tolerance of the inflatable) between corrections applied by the Pressure Correction System. For more information on the design of these systems see Arenberg et al. (2021) \cite{ArenbergArch}.

As an example, let us calculate the gas lost due to the micrometeoroid environment for the astrophysics mission concept OASIS\cite{ArenbergArch22}. OASIS is a one year mission which includes a 14 $m$ diameter inflatable reflector located at the L1 Lagrange point. The lenticular is filled with neon gas at a pressure of 3.5 $Pa$, with typical gas temperatures of $\sim$300 $K$. In this thought experiment, we will focus on the gas lost solely due to micrometeoroid punctures, so we will set $j$=0 in Equation \ref{Eq49}. In Section \ref{sect: mitigation} we found that we can expect a total damage rate due to the full spectrum of micrometeoroids in this region to be 1.3x10$^{-6}$ $\frac{cm^2}{m^2 yr}$ (this is the integral term in Equation \ref{Eq46}). Plugging in values into Equations \ref{Eq46} and \ref{Eq49} gives $m_G (1 yr)$=1.5x10$^{-3}$ $kg$. Note that the actual gas replenishment may be significantly larger than this once the seams term is included. The authors would like to remind the readers that this estimate is based on the $\kappa$ extrapolation using the observed $\beta$ from the experiments; more hypervelocity impact tests are needed to refine the $\beta$ parameter.

In addition to leaks caused by micrometeoroid penetrations, permeation of the pressurant gas through the reflector membrane is a second mechanism for gas loss that must be considered. The permeability of gasses through polymeric material can vary widely as a function of the permeant gas species, the membrane material, and the temperature. The permeation rate is given by Darcy’s Law 
\begin{equation}
    \label{Eq50}
    Q_{permeation}=K \frac{A_{R}}{f} \Delta P
\end{equation}
where $Q_{permeation}$ is the number flow rate, $K$ is the permeability, $A_{R}$ is the surface area of the reflector, $f$ is the membrane thickness, and $\Delta P$ is the pressure differential of the permeant gas across the membrane (here simply the gas pressure within the reflector). The temperature dependence of the permeability is given by
\begin{equation}
    \label{Eq51}
    K \propto exp \left( -\frac{E_{K}}{k_B T} \right)
\end{equation}
where $E_{K}$ is the energy of permeation. The mass loss from the gas permeating through the reflector material over an elapsed time $t$ can then be found by
\begin{equation}
    \label{Eq53}
   m_{permeation}(t)=\rho_g Q t
\end{equation}
where $\rho_g$ is the mass density of the gas, which can be solved using the ideal gas law 
\begin{equation}
    \label{Eq54}
   \rho_g = \frac{PM}{RT}
\end{equation}
with $M$ the molar mass and $R$ the gas constant. Substituting this into Equation \ref{Eq53} we can solve for the total mass lost over the mission lifetime $Z$ as
\begin{equation} 
    \label{Eq_m_permeation}
   m_{permeation}(Z)=\frac{PM}{RT} Q Z.
\end{equation}
Comparing Equations \ref{Eq49}  and \ref{Eq_m_permeation}, we can see that the gas loss due to holes is proportional to $Z^2$ while loss due to permeation is linearly dependent on $Z$. 

To understand which loss mechanism will dominate, let us again examine the case for OASIS. Schowalter et al. (2010) \cite{Schowalter} found $K=$3.1x10$^{-11}$ $\frac{cm^3 at STP mm}{s torr cm^2}$ for neon. Plugging in the relevant numbers gives $Q_{permeation}=$2x10$^{-10}$ $m^3/s$. This results in a $m_{permeation}(1 yr)=$2x10$^{-7}$ $kg$, four orders of magnitude smaller than the micrometeoroid effect. Therefore, gas escape through punctures and seams will be the dominant source of gas loss.

\section{Summary and Future Work}
\label{sect: summary}

Inflatable reflectors present a unique opportunity to enable large aperture telescopes in space. However, the lifetime of their functionality will depend on the system's ability to maintain a constant shape by maintaining the pressure within the structure. In this paper, we have shown how gas loss due to micrometeoroid punctures will be the dominant loss mechanism of this life-limiting resource. We have also presented a theoretical formulation for predicting the hole damage due to a distribution of impacting fragments to better enable estimations of this gas loss rate. 

Hypervelocity tests using a ruby impactor on kapton sheets indicate a micrometeoroid will fragment with an $\alpha_f$=4 and a $c_{max}$=0.24. We use these empirical values to create a predictive model for micrometeoroid impacts on a polyimide film. Using this model, we estimate a micrometeoroid shield consisting of 6-7 layers (depending on the thickness of the polyimide sheets) would be needed to properly mitigate the micrometeoroid environment at the 1 AU location and extend the lifetime of the inflatable optic.

More tests are needed to determine how efficiently the micrometeoroids are eliminated at each membrane barrier, refine the $\beta$ parameter variation with velocity, investigate fragment elongation, and determine effects of stand-off distance between the membranes. We have presented a prediction for future tests which incorporate multiple membranes to replicate the efficacy of a micrometeoroid shield. Comparisons of those tests to predictions can help refine the current model.

\section* {Acknowledgments}
Support for this project provided by Northrop Grumman internal funds. We would like to thank the staff at White Sands Hypervelocity Test Facility. We also would like to thank Tiffany Glassman for helpful comments resulting in a more readable manuscript. A version of Section \ref{sect: gas_loss_deriv} has previously appeared in a SPIE conference proceeding: Proceedings Volume 11820, Astronomical Optics: Design, Manufacture, and Test of Space and Ground Systems III; 118200T (2022) https://doi.org/10.1117/12.2594706.


\bibliography{report}   
\bibliographystyle{spiejour}   


\vspace{2ex}\noindent\textbf{Michaela Villarreal}
 is a Systems Engineer at Northrop Grumman working on mission concept development for astrophysics and planetary space missions. She received her Bachelors degree in Planetary Science from UC Berkeley and her M.S. and Ph.D. degrees from UCLA in Geophysics and Space Physics. She was previously a science team affiliate of NASA's Dawn and Europa Clipper Missions.
\vspace{1ex}

\vspace{2ex}\noindent\textbf{Jonathan Arenberg} is Chief Mission Architect for Science and Robotic Missions at Northrop Grumman Corporation. His degrees are in physics and engineering, from the University of California, Los Angeles.  Dr. Arenberg has contributed to the Chandra X-ray Observatory, Starshade, the James Webb Space Telescope and the Astro 2020 strategic missions.  He is widely published, awarded 15 European and U.S. Patents and is an SPIE Fellow.
\vspace{1ex}

\vspace{2ex}\noindent\textbf{Lauren Halvonik Harris}
 is a Systems Engineer for Northrop Grumman, working in the Radiation Effects and Survivability area. She received her Bachelors degree in physics from UCLA. She was previously an accelerator engineer at RadiaBeam technologies working on the development of plasmonic Niobium photocathodes for SRF gun applications in collaboration with Jefferson Lab.
\vspace{1ex}

\listoffigures
\listoftables

\end{spacing}
\end{document}